\documentclass[prl,twocolumn,superscriptaddress]{revtex4}

\usepackage{graphicx}
\usepackage{amsmath,amsfonts}
\usepackage{bbm}
\usepackage[pdftex,usenames,dvipsnames]{color}

\renewcommand{\eqref}[1]{Eq.~(\ref{#1})}
\newcommand{\ket}[1]{\vert#1\rangle}

\def\opone{\leavevmode\hbox{\small1\kern-3.8pt\normalsize1}}

\begin{document}

\title{Two-photon interference of weak coherent laser pulses recalled from separate solid-state quantum memories}

\author{Jeongwan Jin}
\affiliation{Institute for Quantum Science and Technology, and Department of Physics \& Astronomy, University of Calgary, 2500 University Drive NW, Calgary, Alberta T2N 1N4, Canada}
\author{Joshua A. Slater}
\affiliation{Institute for Quantum Science and Technology, and Department of Physics \& Astronomy, University of Calgary, 2500 University Drive NW, Calgary, Alberta T2N 1N4, Canada}
\author{Erhan Saglamyurek}
\affiliation{Institute for Quantum Science and Technology, and Department of Physics \& Astronomy, University of Calgary, 2500 University Drive NW, Calgary, Alberta T2N 1N4, Canada}
\author{Neil Sinclair}
\affiliation{Institute for Quantum Science and Technology, and Department of Physics \& Astronomy, University of Calgary, 2500 University Drive NW, Calgary, Alberta T2N 1N4, Canada}
\author{Mathew George}
\affiliation{Department of Physics - Applied Physics, University of Paderborn, Warburger Strasse 100, 33095 Paderborn, Germany}
\author{Raimund Ricken}
\affiliation{Department of Physics - Applied Physics, University of Paderborn, Warburger Strasse 100, 33095 Paderborn, Germany}
\author{Daniel Oblak}
\affiliation{Institute for Quantum Science and Technology, and Department of Physics \& Astronomy, University of Calgary, 2500 University Drive NW, Calgary, Alberta T2N 1N4, Canada}
\author{Wolfgang Sohler}
\affiliation{Department of Physics - Applied Physics, University of Paderborn, Warburger Strasse 100, 33095 Paderborn, Germany}
\author{Wolfgang Tittel}
\affiliation{Institute for Quantum Science and Technology, and Department of Physics \& Astronomy, University of Calgary, 2500 University Drive NW, Calgary, Alberta T2N 1N4, Canada}

%Intro: [189/150]
%--------------------------------------------------------------------------------------------
% Abstract
%--------------------------------------------------------------------------------------------
\begin{abstract}
Quantum memories for light, which allow the reversible transfer of quantum states between light and matter, are central to the development of quantum repeaters \cite{sangouard2011a}, quantum networks \cite{kimble2008a}, and linear optics quantum computing \cite{kok2007a}. Significant progress has been reported in recent years, including the faithful transfer of quantum information from photons in pure and entangled qubit states \cite{lvovsky2009a,saglamyurek2011a,clausen2010a,saglamyurek2012a,zhang2011a,specht2011a,england2012a}. 
However, none of these demonstrations confirm that photons stored in and recalled from quantum memories remain suitable for two-photon interference measurements, such as C-NOT gates and Bell-state measurements, which constitute another key ingredient for all aforementioned applications of quantum information processing.
Using pairs of weak laser pulses, each containing less than one photon on average, we  demonstrate two-photon interference as well as a Bell-state measurement after either none, one, or both pulses have been reversibly mapped to separate thulium-doped titanium-indiffused lithium niobate (Ti:Tm:LiNbO$_3$) waveguides. 
As the interference is always near the theoretical maximum, we conclude that our solid-state quantum memories, in addition to faithfully mapping quantum information, also preserves the entire photonic wavefunction.
Hence, we demonstrate that our memories are generally suitable for use in advanced applications of quantum information processing that require two-photon interference.
\end{abstract}

\maketitle

%--------------------------------------------------------------------------------------------
% Main
%--------------------------------------------------------------------------------------------

%Main: [1717/1500]

When two indistinguishable single photons impinge on a 50/50 beam-splitter (BS) from different input ports, they bunch and leave together by the same output port. This so-called Hong-Ou-Mandel (HOM) effect \cite{hong1987a} is due to destructive interference between the probability amplitudes associated with both input photons being transmitted or both reflected, see Fig. 1. Since no such interference occurs for distinguishable input photons, the interference visibility $V$ provides a convenient way to verify that two photons are indistinguishable in all degrees of freedom, i.e. spatial, temporal, spectral, and polarization modes. The visibility is defined as
\begin{equation}
\label{eq:HOM_Visibility}
	V = (\mathcal{R}_\mathrm{max} - \mathcal{R}_\mathrm{min})/\mathcal{R}_\mathrm{max},
\end{equation}
where $\mathcal{R}_\mathrm{min}$ and $\mathcal{R}_\mathrm{max}$ denote the rate with which photons are detected in the two output ports in coincidence if the incoming photons are indistinguishable and distinguishable, respectively.  Consequently, the HOM effect has been employed to characterize the indistinguishability of photons emitted from a variety of sources, including 
parametric down-conversion crystals \cite{kaltenbaek2006a}, trapped neutral atoms \cite{beugnon2006a,specht2011a}, trapped ions \cite{maunz2007a}, quantum dots \cite{sanaka2009a,patel2010a,flagg2010a}, organic molecules \cite{lettow2010a}, nitrogen-vacancy centres in diamond \cite{bernien2011a, sipahigil2012a}, and atomic vapours \cite{felinto2006a,chaneliere2007a,yuan2007a,chen2008a,yuan2008a}. Furthermore, two-photon interference is at the heart of linear optics Bell-state measurements \cite{weinfurter1994a}, and, as such, has already enabled experimental quantum dense coding \cite{mattle1996a}, quantum teleportation \cite{bouwmeester1997a}, and entanglement swapping \cite{pan1998a}. However, to date, the possibility to perform Bell-state measurements with photons that have previously been stored in a quantum memory, as required for advanced applications of quantum information processing, has not yet been established. For these measurements to succeed, photons need to remain indistinguishable in all degrees of freedom, which is more restrictive than the faithful recall of encoded quantum information. Indeed, taking into account that photons may or may not have been stored before the measurement, this criterion amounts to the requirement that a quantum memory preserves a photon's wavefunction during storage. Similar to the case of photon sources, the criterion of indistinguishability is best assessed using HOM interference, provided single-photon detectors are employed. 

\begin{figure*}
	\includegraphics[width=0.95\textwidth]{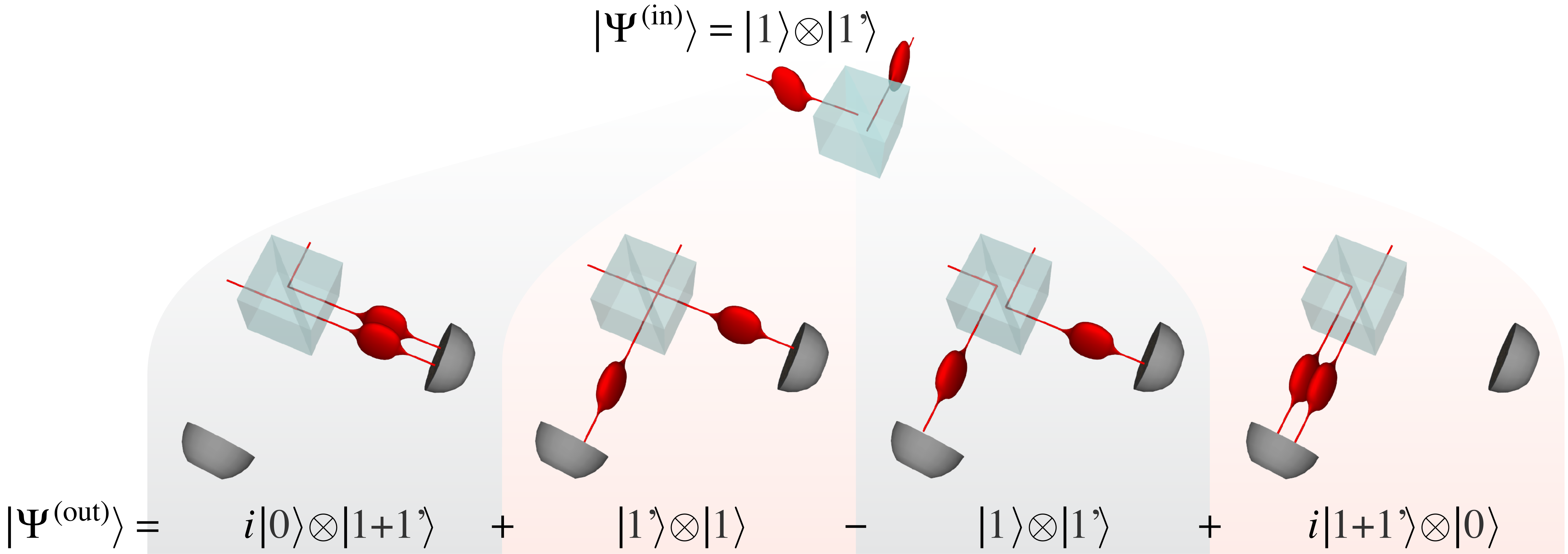}%
	\caption{Illustration of HOM-interference in the case of single photons at BS input $|\psi^\mathrm{(in)}\rangle=|1,1'\rangle$, where the prime on the latter input indicates the possibility to distinguish that input photon from the other in some degree of freedom e.g. by being polarized orthogonally. The four possible paths of the photons are illustrated, together with their corresponding output states. If the input photons are indistinguishable with respect to all degrees of freedom we can ignore the primes in the output states and the paths shown in the two central pictures are identical and, due to the different signs, thus cancel. This leaves in the output state $|\psi^\mathrm{(out)}\rangle$ only the possibilities in which photons bunch. For distinguishable photons, e.g. having orthogonal polarizations, all paths are distinguishable and all terms remain in $|\psi^\mathrm{(out)}\rangle$.}
\label{fig:homillustr}
\end{figure*}

\begin{figure*}[t!]
	\includegraphics[width=\textwidth]{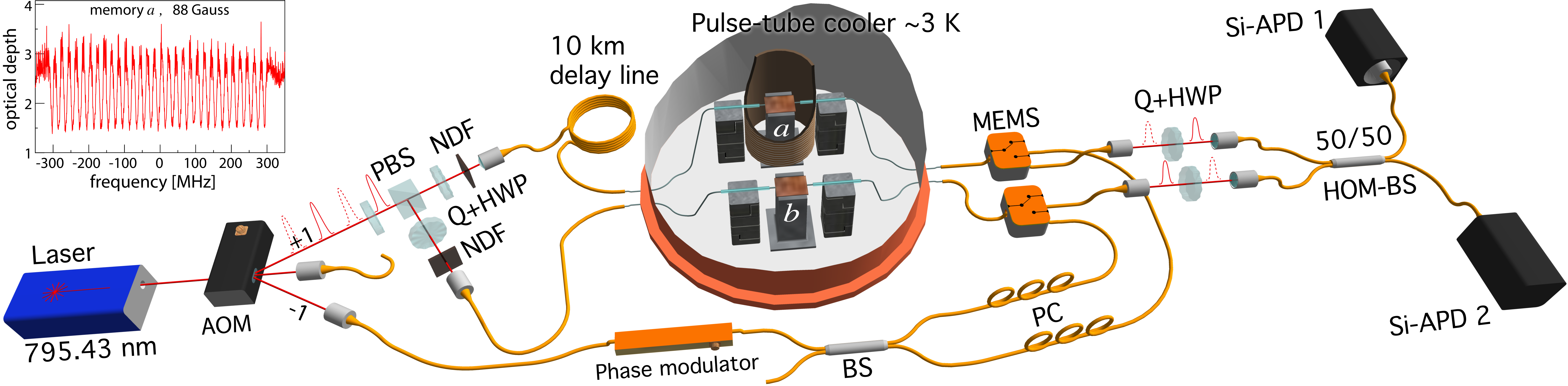}%
	\caption{\textbf{Experimental setup.}
	Light from a 795.43~nm wavelength CW laser passes through an acousto-optic modulator (AOM) driven by a sinusoidally varying signal. The first negative refraction order is fibre coupled into a phase modulator and, via a beam-splitter (BS), two polarization controllers (PCs) and two micro-electromechanical switches (MEMS), injected from the back into two Ti:Tm:LiNbO$_3$ waveguides (labelled $a$ and $b$) cooled to 3 K [\citenum{sinclair2010a}]. Waveguide $a$ is placed inside a superconducting solenoid. Using a linear frequency-chirping technique \cite{reibel2004a} we tailor AFCs with 600~MHz bandwidth and a few tens of MHz peak spacing, depending on the experiment, into the inhomogeneously broadened absorption spectrum of the thulium ions, as shown for crystal $a$ in the inset. After 3~ms memory preparation time and 2~ms wait time we store and recall probe pulses during 3~ms. The 8~ns long probe pulses with $\approx$ 50~MHz Fourier-limited bandwidth are derived from the first positive diffraction order of the AOM  output at a repetition rate of 2.5-3~MHz. Each pulse is divided into two spatial modes by a half-wave plate (HWP) followed by a polarizing beam-splitter (PBS). All pulses are attenuated by neutral-density filters (NDFs) and coupled into optical fibres and injected from the front into the Ti:Tm:LiNbO$_3$ waveguides. After exiting the memories (i.e. either after storage, or after transmission), the pulses pass quarter- and half-wave plates used to control their polarizations at the 50/50 BS (HOM-BS) where the two-photon interference occurs.
Note that, to avoid first-order interference, pulses passing through memory $a$ propagate through a 10~km fibre to delay them w.r.t. the pulses passing through memory $b$ by more than the laser coherence length. Finally, they are detected by two single-photon detectors (actively quenched silicon avalanche photodiodes, Si-APDs) placed at the outputs of the beam-splitter, and coincidence detection events are analyzed with a time-to-digital convertor (TDC) and a computer.}
\label{fig:experimentalsetup}
\end{figure*}

Our experimental setup is depicted in Fig.~\ref{fig:experimentalsetup}. We employ solid-state quantum memories, more precisely thulium-doped lithium-niobate waveguides in conjunction with the atomic frequency comb (AFC) quantum memory protocol \cite{afzelius2009a}, which have shown great promise for advanced applications of quantum information processing \cite{saglamyurek2011a,clausen2010a,saglamyurek2012a}. We then interfere various combinations of recalled and non-stored (i.e. directly transmitted) pulses on a 50/50 BS (HOM-BS). When using single photon Fock states at the memory inputs, the HOM visibility given in \eqref{eq:HOM_Visibility} theoretically reaches 100\% as illustrated in Fig.~\ref{fig:homillustr}. However, with phase incoherent laser pulses obeying Poissonian photon-number statistics, as in our demonstration, the maximally achievable visibility is 50\% \cite{mandel1983a}, irrespective of the mean photon number (see Supplementary Information). Nevertheless, attenuated laser pulses are perfectly suitable for assessing the effect of our quantum memories on the photonic wavefunction. Any reduction of indistinguishability due to storage causes a reduction of visibility, albeit from maximally 50\%. This approach extends the characterization of quantum memories using attenuated laser pulses \cite{riedmatten2008a} from assessing the preservation of quantum information during storage to assessing the preservation of the entire wavefunction, and from first- to second-order interference.

%\section*{Single-photon-level storage and modification of the  polarization degree of freedom}

We first deactivate both quantum memories (see Supplementary Information), to examine the interference between directly transmitted pulses, and thereby establish a reference visibility for our experimental setup. We set the mean photon number per pulse before the memories to 0.6, i.e. to the single-photon level.
Using the wave plates, we rotate the polarizations of the pulses at the two HOM-BS inputs to be parallel (indistinguishable) or orthogonal (distinguishable). Employing \eqref{eq:HOM_Visibility} we find a visibility of $(47.9\pm 3.1)$\%.

Subsequently, we activate memory $a$ while keeping memory $b$ off, and adjust the timing of the pulse preparation so as to interfere a recalled pulse from the active memory with a directly transmitted pulse from the inactive memory (see Supplementary Information).
Pulses are stored for 30~ns in memory $a$, and the mean photon number per pulse at the quantum memory input is 0.6. Taking the limited storage efficiency of $\approx1.5\%$ and coupling loss into account, this results in $3.4\times10^{-4}$ photons per pulse at the HOM-BS inputs. As before, changing the pulse polarizations from mutually parallel to orthogonal, we find $V=(47.7\pm 5.4)$\%, which equals our reference value within the measurement uncertainties.

As the final step, we activate both memories to test the feasibility of two-photon interference in a quantum-repeater scenario. We note that in a real-world implementation, memories belonging to different network nodes are not necessarily identical in terms of material properties and environment. This is captured by our setup where the two Ti:Tm:LiNbO$_3$ waveguides feature different optical depths and experience different magnetic fields (see Fig.~\ref{fig:experimentalsetup} and Supplementary Information). To balance the ensuing difference in memory efficiency we set the mean photon number per pulse before the less efficient and more efficient memories to 4.6 and 0.6, respectively, so that, as before, the mean photon numbers are $3.4\times10^{-4}$ at both HOM-BS inputs. With the storage time of both memories set to 30~ns, we get $V=(47.2\pm 3.4)$\%, in excellent agreement with the values from the previous measurements. The consistently high visibilities, compiled in the first column of Table~\ref{tab:visresults}, hence confirm that our storage devices do not introduce any degradation of photon indistinguishability during the reversible mapping process, and that two-photon interference is feasible with photons recalled from separate quantum memories, even if the memories are different.

%\section*{Storage of few-photon pulses and modification of all available degrees of freedom}

We now investigate in greater detail the change in coincidence count rates as photons gradually change from being mutually indistinguishable to completely distinguishable w.r.t.~each degree of freedom accessible for change in single-mode fibres, i.e.  polarization, temporal, and spectral modes (see Supplementary Information). To acquire data more efficiently we increase the mean number of photons per pulse at the memory input to between 10 and 50 (referred to as few-photon-level measurements). However, the mean photon number at the HOM-BS remains below one. Example data plots are shown in Fig.~\ref{fig:exresults}, while the complete set of plots is supplied in the Supplementary Information Figs.~\ref{fig:coin-0qm}-\ref{fig:coin-2qm}.

In Fig.~\ref{fig:exresults}a we show the coincidence counts rates as a function of the polarization of the recalled pulse for the case of one active memory. The visibilities for all configurations (i.e. zero, one, or two active memories) extracted from fits to the experimental data are listed in column 2 of Table~\ref{tab:visresults}. They are -- as in the case of single-photon-level inputs -- equal to within the experimental uncertainty.

\begin{figure*}[t!]
	\centering
	\includegraphics[width=\textwidth]{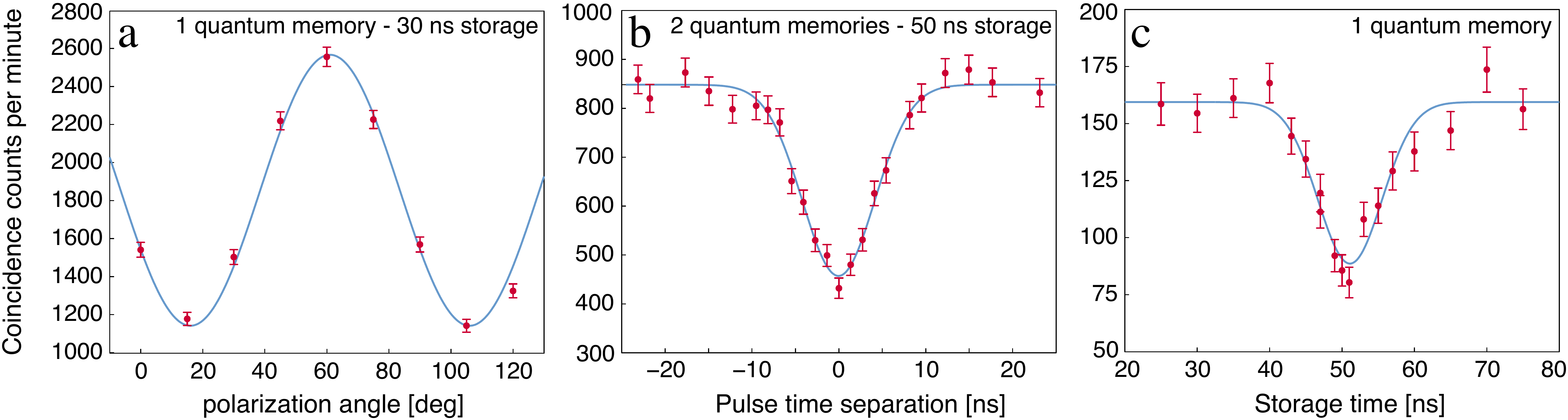}
	\caption{HOM interference plot examples for one or two active memory configurations (as labelled). a) Varying mutual polarization difference. b) Varying temporal overlap by changing timing of pulse generation. c) Varying temporal overlap by changing storage time. The acquisition time per data point is 60~s in a,b and 120~s in c.}
	\label{fig:exresults}
\end{figure*}

Next, in Fig.~\ref{fig:exresults}b, we depict the coincidence count rates as a function of the temporal overlap (adjusted by the timing of the pulse generation) for the two-memory configuration. Column 3 of Table~\ref{tab:visresults} shows the visibilities extracted from Gaussian fits to the data, reflecting the temporal profiles of the probe pulses, for all configurations. Within experimental uncertainty, they are equal to each other. 
Alternatively, in the single-memory configuration, we also change the temporal mode overlap by adjusting the storage time of the pulse mapped to the quantum memory. Again the measured visibility of $V=(44.4\pm6.9)$\% (see Fig.~\ref{fig:exresults}c) is close to the theoretical maximum.

Finally, we vary the frequency difference between the two pulses (see Supplementary Information) to witness two-photon interference w.r.t.~spectral distinguishability. For this measurement, we consider only the configurations in which neither, or a just single memory is active. In both cases the visibilities, listed in the last column of Table~\ref{tab:visresults}, are around 43\%. While this is below the visibilities found previously, for reasons discussed in the Supplementary Information, the key observation is that the quantum memory does not affect the visibility. 

\begin{table}[h]
	\caption{Experimental two-photon interference visibilities (\%) for different degrees of freedom}
	\centering
	\begin{tabular}{c | c | c c c}
		\hline\hline
		 & Single-photon & \multicolumn{3}{c}{Few-photon} \\
		Storage & level & \multicolumn{3}{c}{level} \\[1mm]
		\cline{2-5}\\
		configuration & Polarization & \multicolumn{1}{c}{ Polarization} &  \multicolumn{1}{c}{Temporal} & Spectral\\
		\hline\hline
		No-storage & $47.9\pm 3.1$ & $51.0\pm5.6$ & $42.4\pm 2.3$ & $43.7\pm1.7$ \\ 
		Single-storage & $47.7\pm 5.4$ & $55.5\pm 4.1$ & $47.6\pm 3.0$ & $42.4\pm3.5$ \\ 
		Double-storage& $47.2\pm 3.4$ & $53.1\pm 5.3$ & $46.1\pm 3.2$ & N. A. \\[1mm]
		\hline
	\end{tabular}
	\label{tab:visresults}
\end{table}

%\section*{Bell-state measurement}

As stated in the introduction, Bell-state measurements (BSM) with photonic qubits recalled from separate quantum memories are key ingredients for advanced applications of quantum communication. To demonstrate this important element, we consider the asymmetric (and arguably least favourable) case in which only one of the qubits is stored and recalled.
Appropriately driving the AOM in Fig.~\ref{fig:experimentalsetup}, we prepare the states $|\Psi_1\rangle$ and $|\Psi_2\rangle$, which describe time-bin qubits \cite{tittel2001a} of the form $|e\rangle$, $|l\rangle$, $\frac{1}{\sqrt{2}}(\ket{e}+\ket{l})$, or $\frac{1}{\sqrt{2}}(\ket{e}-\ket{l})$, where $e$ and $l$, respectively, label photons in early or late temporal modes, which are separated by 25~ns. The qubits are directed to the memories of which only one is activated. The mean photon number of the qubit that is stored is set to 0.6, yielding a mean photon number of both qubits at the HOM-BS input of $6.7\times10^{-4}$. We ensure to overlap pulses encoding the states $|\Psi_1\rangle$ and $|\Psi_2\rangle$ at the HOM-BS and count coincidence detections that correspond to a projection onto the $\ket{\psi^-}=\frac{1}{\sqrt{2}}(\ket{e}\ket{l}-\ket{l}\ket{e})$ Bell state.
This projection occurs if the two detectors click with 25~ns time difference \cite{tittel2001a}. Because $\ket{\psi^-}$ is antisymmetric w.r.t. any basis, the count rate is expected to reach a minimum value $\mathcal{R}_{\parallel}$ if the two input pulses are prepared in equal states, and a maximum value $\mathcal{R}_{\bot}$ if prepared in orthogonal states.
Accordingly, we define an error rate that quantifies the deviation of the minimum count rate from its ideal value of zero:
\begin{align}\label{eq:errdef}
	e\equiv\frac{\mathcal{R}^{\parallel}}{\mathcal{R}^{\parallel}+\mathcal{R}^{\perp}} \, .
\end{align}
First, choosing to encode  $|\Psi_1\rangle$ and $|\Psi_2\rangle$ in states $|e\rangle$ and $|l\rangle$ we obtain the error rate $e_{e/l}^{(\mathrm{exp})}=0.039\pm0.037$, which is near the theoretical value of $e_{e/l}^{(\mathrm{QM})}=0$ (see the Supplementary Information for derivations of the theoretical values and bounds). In addition it clearly violates the lower bound $e_{e/l}^{(\mathrm{CM})}=0.33$ that can be obtained for a Bell-state measurement on two qubits of which one is recalled from a classical memory (CM). Note that values for $e_{e/l}^{(\mathrm{QM})}$ and $e_{e/l}^{(\mathrm{CM})}$ are independent of whether $|e\rangle$ and $|l\rangle$ qubits are encoded into single photons or attenuated laser pulses. Next, using instead the states $|+\rangle\equiv\frac{1}{\sqrt{2}}(\ket{e}+\ket{l})$, and $|-\rangle\equiv \frac{1}{\sqrt{2}}(\ket{e}-\ket{l})$ we measure $e_{+/-}^{(\mathrm{exp})}=0.287\pm0.020$, which again only slightly exceeds the lowest possible value for attenuated laser pulses of $e_{+/-}^{(\mathrm{att,QM})}=0.25$. 
The crucial observation is once more that $e_{+/-}^{(\mathrm{exp})}$ violates both the lower bound for qubits encoded into single photons $e_{+/-}^{(\mathrm{sing,CM})}=0.33$ and attenuated laser pulses $e_{+/-}^{(\mathrm{att,CM})}=0.417$ that are recalled from a classical memory.

%\section*{Conclusion}

Our demonstrations show that solid-state AFC quantum memories are suitable for two-photon interference experiments, even in the general case of storing the two photons an unequal number of times. With improved system efficiency \cite{afzelius2010b} and multi-mode storage supplemented by read-out on demand%
 \cite{sinclair2012a,afzelius2010a,gundogan2013a}, such memories can be used as synchronization devices in multi-photon experiments, which will allow increasing the number of photons that can be harnessed simultaneously for quantum information processing beyond the current limit of eight \cite{yao2012a}. A subsequent goal is to develop workable quantum repeaters or, more generally, quantum networks, for which longer storage times are additionally needed. Depending on the required value, which may range from hundred micro-seconds \cite{munro2010a} to seconds \cite{sangouard2011a,timoney2013a}, this may be achieved by storing quantum information in optical coherence, or it may require mapping of optical coherence onto spin states \cite{afzelius2009a}.

\newpage

\section{Supplementary Information}

\subsection{Memory operation and properties.}
A quantum memory is said to be activated when we configure the MEMS to allow the optical pumping light to reach the waveguide during the preparation stage and thus tailor an AFC in the inhomogeneously broadend absorption spectrum of thulium ions (see Fig.~\ref{fig:experimentalsetup}). If the optical pumping is blocked, the memory is said to be deactivated and light entering the waveguide merely experiences constant attenuation over its entire spectrum.
If a memory is activated, an incident photon is mapped onto a collective excitation of thulium ions in the prepared AFC and subsequently re-emitted at a time given by the inverse of the comb tooth spacing \cite{afzelius2009a}, i.e., $t=1/\Delta$ (see Fig.~\ref{fig:experimentalsetup}). 
In all cases, we adjust the mean photon number at the memory inputs so that mean photon numbers are equal at the HOM-BS inputs. This is required for achieving maximum visibility with attenuated laser pulses (further details later in Supplementary Information).

The two Ti:Tm:LiNbO$_3$ waveguides are fabricated identically but differ in terms of overall length,
yielding optical depths of 2.5 for memory $a$ and 3.2 for memory $b$. 
As shown in Figure 1, memory $a$ is placed at the centre of a solenoid in a uniform magnetic field, while memory $b$ is placed outside the solenoid and thus experiences only a much weaker stray field. Therefore it is not possible to achieve the optimal efficiency for both memories at the same time (further details later in Supplementary Information).

\subsection{Changing degrees of freedom.}
\textbf{a)} The polarization degree is easily adjusted using the free-space half- and quarter-wave plate set at each HOM-BS input. For our measurements we rotate the half-wave plate in steps of either 45 or 7.5 degrees. \textbf{b)} The temporal separation $\delta t$ between a pulse arriving at one of the HOM-BS inputs and the next pulse in the train arriving at the other input can be expressed as $\delta t =  \{nl/c \}\, \mathrm{mod}~\delta t_r$, where $n$ is the refractive index of the fibres, $l\approx10$~km is the path-length difference for pulses interacting with memory $a$ and $b$, and $\delta t_r$ is the repetition period of the pulse train from the AOM, which is set in the range of 350-400~ns. As we can change $\delta t_r$ with 10~ps precision, we can tune $\delta t$ on the ns scale. \textbf{c)} For the storage time scan, the recall efficiency decreases with storage time due to decoherence. Hence, we balance the mean photon number per pulse for stored and transmitted pulses for each storage time. \textbf{d)} Finally, to change the spectral overlap of the pulses input to the HOM-BS we can utilize that these pulses were generated at different times in the AOM and thus we can chose their carrier frequencies independently. We interchangeably drive the AOM by frequencies $\nu_a$ and $\nu_b$ and thus create two interlaced trains of pulses with different frequencies. Adjusting the pulse timing we can ensure that the pulses overlapped at the HOM-BS belong to different trains and thus have a spectral overlap given by $\delta\nu=\nu_a-\nu_b$. Due to the limited bandwidth of the AOM we are only able to scan $\delta\nu$ by 100~MHz, which, when compared to the 50~MHz pulse bandwidth, is not quite sufficient to make the pulses completely distinguishable. To achieve complete distinguishability, we supplement with a measurement using orthogonal polarizations at the inputs (further details later in Supplementary Information). 

\subsection{Preparing states for Bell-state measurement.}
For the Bell-state projection measurement we interchangeably prepare the time-bin qubits in either $|e\rangle$ or $|l\rangle$, or in $\frac{1}{\sqrt{2}}(\ket{e}+\ket{l})$ and $\frac{1}{\sqrt{2}}(\ket{e}-\ket{l})$ by setting the relative phase and intensity of the AOM drive signal. Adjusting the timing of the pulse preparation we ensure that qubits in different states overlap at the HOM-BS.

%\newpage

\subsection{Properties of waveguide LiNbO$_3$ crystal and AFC.}

\begin{figure*}[t!]
	\includegraphics[width=\textwidth]{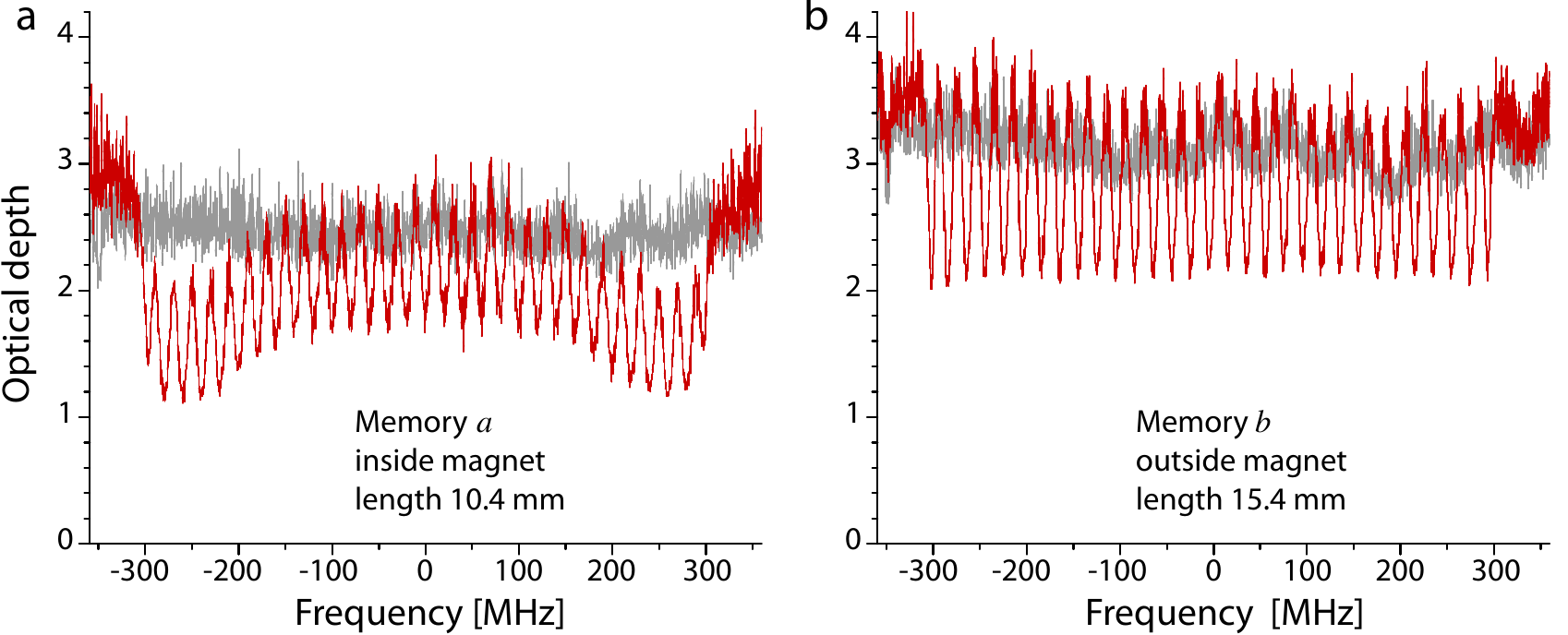}
	\caption{Measured optical depths of our two Ti:Tm:LiNbO$_3$ waveguides as a function of frequency shift of the probing light imparted by the phase-modulator. Light grey traces show optical depths when the memories are inactive, i.e. no AFC is prepared. Dark red traces show the prepared AFCs at a magnetic field of 900~Gauss at the centre of the solenoid.}
	\label{fig:afc-od}
\end{figure*}
\begin{figure*}[t]
	\centering
	\includegraphics[width=0.90\textwidth]{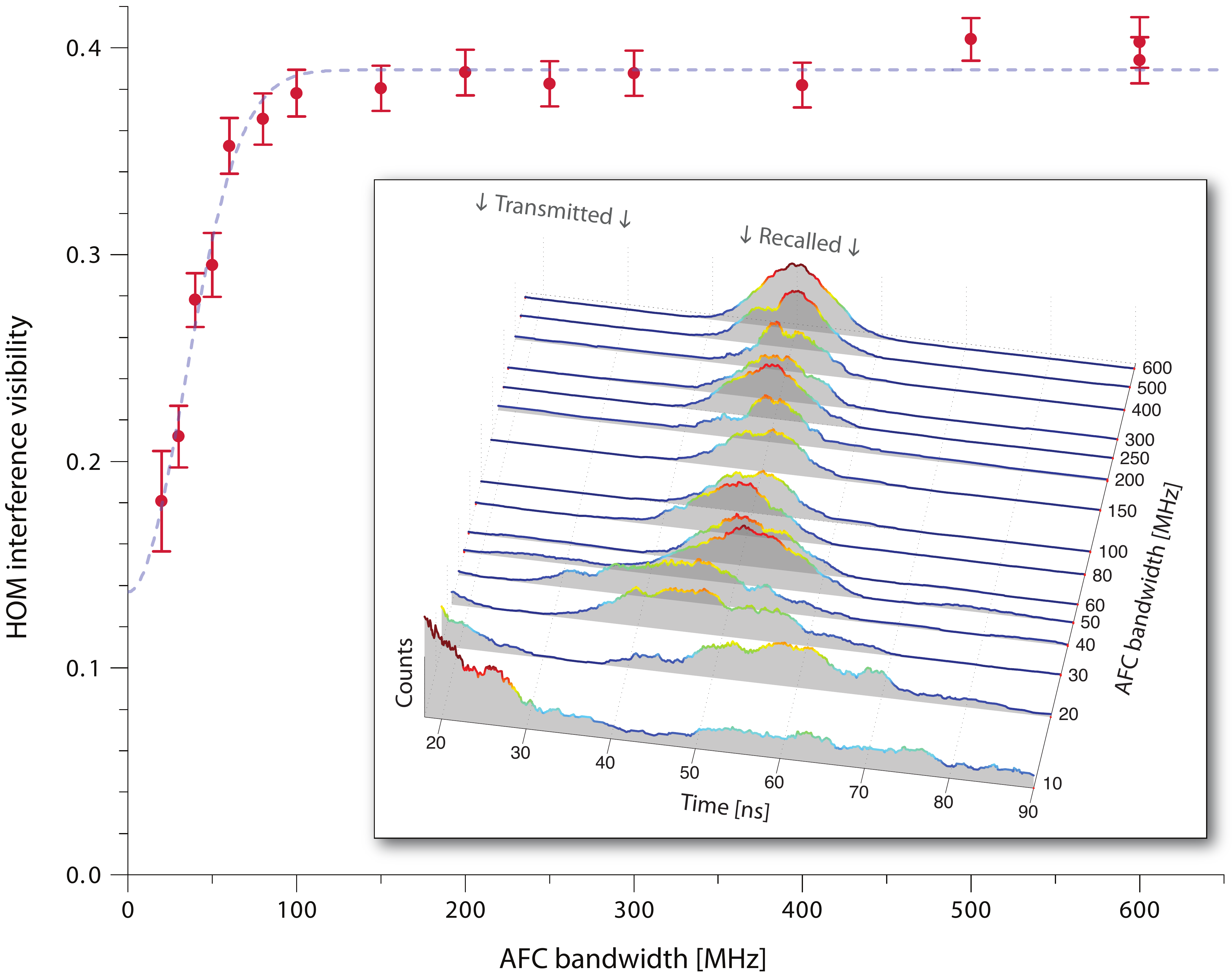}
	\caption{HOM interference visibility if HOM-BS input pulses are recalled from AFCs with varying bandwidths. Insert: Histograms of recalled pulse detection times for different AFC bandwidths clearly showing broadening of recalled (and transmitted) pulses for bandwidths below 100~ns.}
	\label{fig:coin-1qm-bwscan}
\end{figure*}
In the experimental configuration in which the HOM-interference occurs between two pulses recalled from separate quantum memories we pointed, in the main text, to the different properties of the two memory devices. In this section we wish to elaborate on the differences between the two memories based on their physical dissimilarity and measured optical depth as a function of frequency. Memory waveguide $a$ is 10.4~mm long and crystal $b$ is 15.4~mm long. The optical depths at 795.43~nm are around 2.5 and 3.2 for waveguide $a$ and $b$, respectively, as shown by the light-grey curves in Fig.~\ref{fig:afc-od}~a,b, corresponding to the case in which the memories are not activated.

In order to spectrally tailor an AFC in Tm:LiNbO$_3$, a magnetic field must be applied along the crystal's c-axis so as to split the ground and excited level multiplets into their two nuclear Zeeman sublevels \cite{sinclair2010a}. 
However, as one crystal is located at the centre of the setup's solenoid and the other outside the solenoid (see Fig.~\ref{fig:experimentalsetup}) it is not possible to apply the same B-field at the two crystals.
Thus when we activate both memories we generally apply a magnetic field, which provides a reasonable balance in recall efficiencies but is not optimal for either memory. This circumstance is reflected by the different shapes of optical-depth profiles of the AFCs shown in red in Fig.~\ref{fig:afc-od}a,b.

\subsection{Two-photon interference in imperfectly prepared memories.}

In all our demonstrations of the HOM interference we consistently observe that the HOM visibility is close to the theoretical maximum for coherent states. Yet, it is important to realize that an improperly configured AFC quantum memory does alter a stored photon's wavefunction, resulting in imperfect HOM interference with a non-stored photon. 

To support this claim we activate only memory $a$, whose performance we change by varying the bandwidth of the AFC, and interfere the recalled pulses with pulses directly transmitted through the deactivated memory $b$. As the AFC bandwidth decreases below that of the probe pulses, the AFC effectively acts as a bandpass filter for the stored photons and we thus expect the recalled pulses to be temporally broadened w.r.t. the original pulse. This is observed in the insert of Fig.~\ref{fig:coin-1qm-bwscan}, which shows smoothed histograms of photon detection events as a function time. It is worth noting that the small bandwidth AFC also acts as a bandpass filter for the transmitted pulse by virtue of the different effective optical depths inside and outside the AFC. Thus the broadened transmitted pulse starts to overlap with the echo for the narrow AFC bandwidth traces, as is also observed in the insert of Fig.~\ref{fig:coin-1qm-bwscan}.

Another consequence of reducing the AFC bandwidth is that the overall efficiency of the quantum memory decreases, which causes an imbalance between the mean photon numbers at the HOM-BS inputs and thus reduces HOM interference visibility. We circumvent the change to the echo efficiency by adapting the mean photon number at the memory input so as to keep the mean photon number of the recalled pulse constant. With this remedial procedure, we assess the HOM visibility by changing the HOM-BS inputs from parallel to orthogonal polarizations for a series of different AFC bandwidths. The HOM visibility in Fig.~\ref{fig:coin-1qm-bwscan} is steady for bandwidths from around 100~MHz and up. However, below 100~MHz the visibility begins to drop significantly. The dashed line is a fit of the visibilities to a Gaussian function with full-width at half-maximum (FWHM) of $79\pm4~MHz$. Note, that the reason for the visibility being limited to around 40\% is solely that, for this measurement, we do not go through the usual careful optimization steps.

With these measurements we have illustrated how a quantum memory could alter the photonic wavefunction resulting in a reduced HOM interference visibility. A combination of spectral and temporal distortion of the photonic wavefunction is indeed a common type of perturbation by quantum memories. \cite{chaneliere2005a,eisaman2005a} It is particularly worth noting that the gradient-echo memory (GEM) quantum memory protocol, though similar to the AFC protocol, imparts a frequency chirp to the recalled pulse \cite{moiseev2008a}. If not corrected, this feature constitutes a perturbation of the wavefunction of the recalled pulse, which may render it unsuitable for applications relying on two-photon interference.

\subsection{Analytical model of second-order interference in coincidence measurements.}

In the following theoretical treatment we will derive expressions for the coincidence and single-detector counts in terms of probabilities.
By multiplying these probabilities with the average experimental repetition rate we can easily calculate the predicted experimental count rates. 
To a large extent though, we will mainly be interested in relative probabilities or count rates between different settings of the degrees of freedom of pulses. 

It is reasonably straightforward to derive the rates of detection of photons at the outputs of a BS (note that in this Supplementary Information, the HOM-BS of the main text will be referred to as just BS)
In our case coherent states $|\alpha\rangle$ and $|\beta\rangle$, characterized by mean photon numbers $\langle \hat{a}^{\dagger}\hat{a} \rangle = |\alpha|^2$ and $\langle \hat{b}^\dagger \hat{b} \rangle = |\beta|^2$, occupy the two spatial input modes of the BS. In the Fock-basis the coherent state can be represented as 
\begin{equation}\label{eq:cohfockparam}
	|\alpha\rangle = \sum_{n=0}^\infty e^{-\frac{|\alpha|^2}{2}} \frac{\alpha^n}{\sqrt{n!}} |n\rangle = \sum_{n=0}^\infty e^{-\frac{|\alpha|^2}{2}} \frac{\alpha^n}{n!} (\hat{a}^\dagger)^n |0\rangle , 
\end{equation}
and similarly for $|\beta\rangle$.

To account for the cases of photons being distinguishable and indistinguishable at the BS we must allow for an additional degree of freedom in each of the spatial modes, e.g. polarization, frequency, or time. Thus we write the input state at one of the BS inputs as $|\alpha_1,\,\alpha_2 \rangle \equiv |\alpha_1\rangle \otimes |\alpha_2 \rangle$, where $\alpha_1$ and $\alpha_2$ are the coherent state amplitudes in the two orthogonal modes of the auxiliary degree of freedom within the same spatial mode. We treat the coherent state at the other BS input in a similar way.

For the case in which the fields at the inputs of the BS are distinguishable with respect to the auxiliary degree of freedom, the inputs to the BS are described as being in the state $|\alpha ,\, 0 \rangle |0 ,\, \beta \rangle \equiv |\alpha ,\, 0 \rangle \otimes |0 ,\, \beta \rangle$, whereas in the case of them being indistinguishable (up to a difference in the mean photon number) the input fields are written as $|\alpha ,\, 0 \rangle |\beta ,\, 0 \rangle$.

The BS is characterized by its reflection amplitude $r$ and transmission amplitude $t=\sqrt{1-|r|^2}$, which cause the input creation operators to transform as $\hat{a}^\dagger \rightarrow t\hat{c}^\dagger + i r\hat{d}^\dagger$ and $\hat{b}^\dagger \rightarrow i r\hat{c}^\dagger + t\hat{d}^\dagger$. With this in hand, we can compute the state in the BS outputs for any combination of Fock states at the inputs. When the two input states are indistinguishable, i.e. in the same auxiliary degree of freedom, we get \cite{rarity2005a}
\begin{widetext}
\begin{align}\label{eq:fockbsoutind}
	|n,0\rangle|m,0\rangle \rightarrow & \sum_{j=0}^n \sum_{k=0}^m K_{\parallel}(n,m,j,k) \, |j+k ,\, 0 \rangle |n+m-j-k ,\, 0\rangle\\
	& K_{\parallel}(n,m,j,k) = t^{m-k+j} (ir)^{n-j+k} \sqrt{\binom{n}{j} \binom{m}{k} \binom{j+k}{j} \binom{n+m-j-k}{n-j}} , \nonumber
\end{align}
where the binomial coefficient $\binom{x}{y}=\frac{x!}{y!(x-y)!}$. For distinguishable input fields the output state is slightly simpler
\begin{align}\label{eq:fockbsoutdis}
	|n,0\rangle|0,m\rangle \rightarrow & \sum_{j=0}^n \sum_{k=0}^m K_{\perp}(n,m,j,k) \, |j,\,k \rangle |n-j,\,m-k\rangle\\
	&K_{\perp}(n,m,j,k) = \sum_{j=0}^n \sum_{k=0}^m t^{m-k+j} (ir)^{n-j+k} \sqrt{\binom{j}{k} \binom{n-j}{m-k}} . \qquad \qquad \qquad \nonumber
\end{align}
\end{widetext}

The above calculated output modes impinge on the single photon detectors (SPDs). These may be characterized by the probability of detecting an incident single photon. From this single photon detection probability $\eta$ it is also possible to deduce the probability of detecting a pulse consisting of multiple photons, keeping in mind that, irrespective of the number of photons, only a single detection event can be generated. We write $p_1(n)$ for the probability for generating one detector event given $n$ incident photons, and it is useful to note that it relates to the probability $p_0(n)$ of detecting nothing as $p_1(n)=1-p_0(n)$. The probability for not detecting $n$ photons is, on the other hand, easily computed as $p_0(n)=(1-\eta)^n$.
 Since the two detectors at the BS outputs are independent, the probability $p_{11}(n,m)$ of generating a coincidence event, i.e. having simultaneous detection events in each of the detectors, given $n$ and $m$ photons in one and the other output is simply $p_{11}(n,m)=p_1(n)p_1(m)$. Thus the probability for a coincidence detection becomes
\begin{align}\label{eq:nmphotdeteffcoin}
	p_{11}(n,m)&=\left[1-(1-\eta_1)^n \right] \left[1-(1-\eta_2)^m \right] \, ,
\end{align}
where $\eta_1$ and $\eta_2$ are the single photon detection probabilities for detector 1 and 2, respectively. Expressing the coincidence detection probability in terms of Fock states at the BS input we have
\begin{widetext}
\begin{align}\label{eq:nmphotbsoutdetprobcoin}
	P_{11}^{\parallel}(n,m) &= \sum_{j=0}^n \sum_{k=0}^m |K_{\parallel}(n,m,j,k)|^2 \,\, p_{11}(j+k,\, n+m-j-k)\\
	&= \sum_{j=0}^n \sum_{k=0}^m |K_{\parallel}(n,m,j,k)|^2 \, \left[1-(1-\eta_1)^{j+k} \right] \left[1-(1-\eta_2)^{n+m-j-k} \right] \ , \nonumber
\end{align}
\end{widetext}
where $K_{\parallel}(n,m,j,k)$ should be substituted with the factor from \eqref{eq:fockbsoutind}. For distinguishable inputs we find a similar expression for $P_{11}^{\perp}(n,m)$ using the factor $K_{\perp}(n,m,j,k)$ from \eqref{eq:fockbsoutdis}. It is assumed that the detector at a given spatial output mode is equally sensitive to photons in both auxiliary modes, i.e. it detects the states $|k ,\, j \rangle $ and $ |j ,\, k \rangle$ with equal probability.

We are now in the position to formulate an expression for the different detection probabilities given a particular set of coherent input fields. The probability to generate a detection event in both detectors, given coherent input fields of amplitudes $\alpha$ and $\beta$, is
\begin{align}\label{eq:cohphotbsoutdetprobcoin}
	\mathcal{P}^{\parallel (\perp)}_{11}(\alpha,\beta)= \sum_{n=0}^\infty \sum_{m=0}^\infty e^{-|\alpha|^2-|\beta|^2} \frac{(\alpha^{n}\beta^{m})^2}{n!\, m!} P_{11}^{\parallel (\perp)}(n,m) \, .
\end{align}
(Note that to distinguish the probability in \eqref{eq:nmphotbsoutdetprobcoin}, which is applicable to Fock states, from that in\eqref{eq:cohphotbsoutdetprobcoin}, which applies to coherent state inputs, we use $P$ to denote the former and $\mathcal{P}$ for the latter.) This allows us to derive the visibility of the HOM interference on the two detectors as
\begin{align}\label{eq:cohphotcoinvis}
	\mathcal{V}_\mathrm{11}(\alpha,\beta,\eta_1,\eta_2,r)=\frac{\mathcal{P}^{\perp}_{11}(\alpha,\beta)-\mathcal{P}^{\parallel}_{11}(\alpha,\beta)}{\mathcal{P}^{\perp}_{11}(\alpha,\beta)} \, ,
\end{align}
where we have spelled out the parameters that affect the value of the visibility. The quantity $\mathcal{V}_\mathrm{11}$ is referred to as the \emph{HOM visibility}.

\subsection{Simplified model for HOM visibility.}

\begin{figure*}[t!]
	\includegraphics[width=\textwidth]{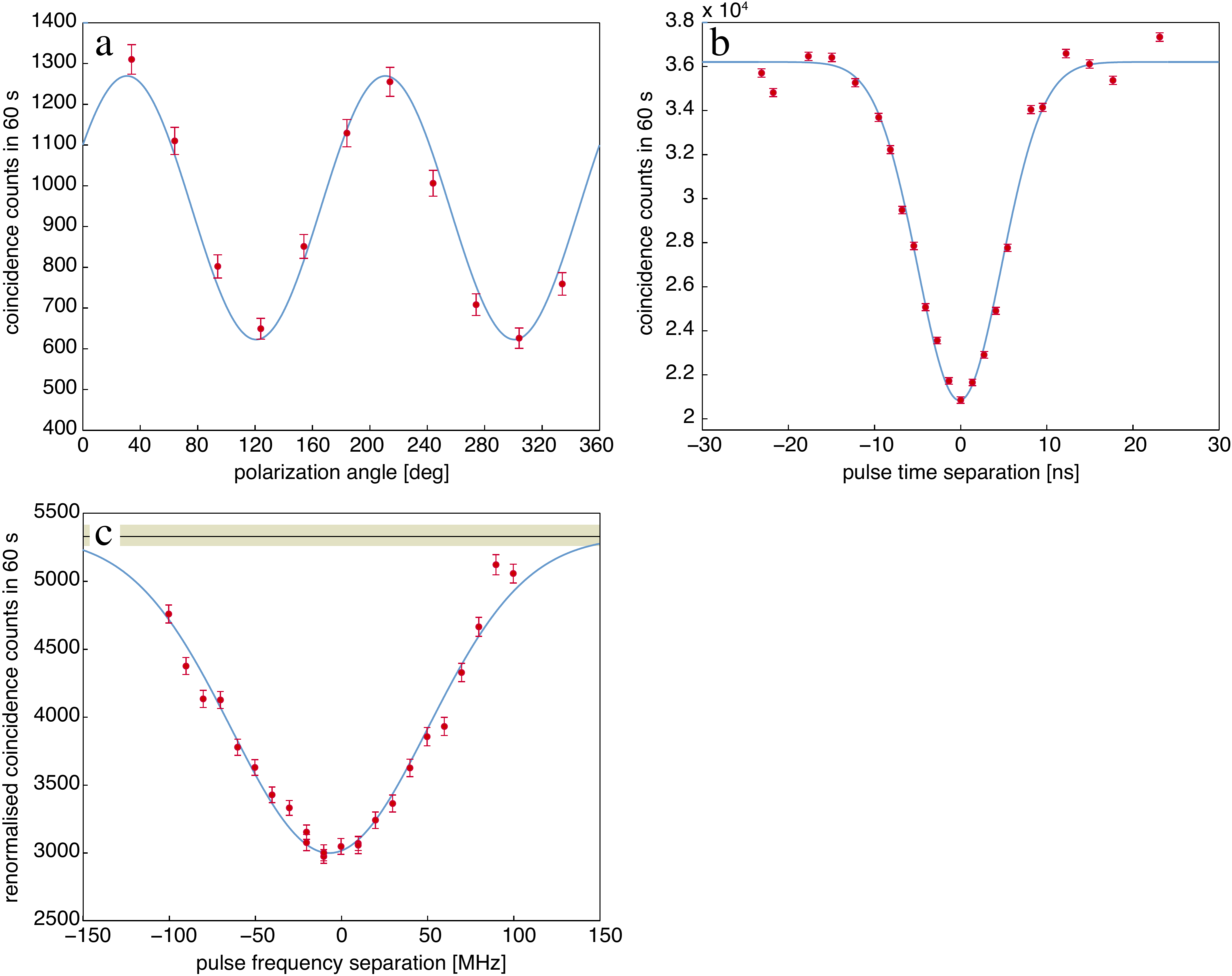}
	\caption{HOM interference manifested in coincidence counts between BS outputs with inactive memories. a) Changing the polarization angle between the pulses yields a HOM visibility of $\mathcal{V}=50.96\pm5.56$\%. b) Varying the temporal overlap of the pulses produces $\mathcal{V}=42.43\pm2.27$\%. c) Altering the frequency overlap of the pulse spectra results in $\mathcal{V}=43.72\pm1.70$\%.}
	\label{fig:coin-0qm}
\end{figure*}

To gain some intuitive understanding of the way the HOM visibility is affected by the experimental parameters we resort to a couple of approximations. Firstly, we assume equal mean photon numbers at the inputs of the beam-splitter, $|\alpha|^2=|\beta|^2\equiv\mu$, the BS ratio to be 50:50 (i.e. $r=t=1/\sqrt{2}$), and the detectors to have equal single photon detection probability $\eta_1=\eta_2\equiv\eta$. Secondly, since we normally work at very low mean photon numbers $\mu < 1$ only the first couple of terms of \eqref{eq:cohfockparam} need to be included. Specifically, we Taylor expand $e^{-\mu/2}$ and keep only terms in the sum up to 2$^\mathrm{nd}$ order in $\mu$. Thus, for the coincidence detection events we get the probabilities
\begin{align}
	\label{eq:cohphotbsoutdetprobcoinsimp}
		\mathcal{P}^{\parallel}_{11} & = \eta^2\frac{\mu^2}{2}\\ 
		\mathcal{P}^{\perp}_{11} & = \eta^2\mu^2 \, ,
\end{align}
which results in a HOM visibility of 
\begin{align}
	\mathcal{V}_\mathrm{11}=\frac{1}{2} \, .
\end{align}
A key point is that the HOM visibility of $50\%$ is independent of the mean photon number $\mu$. This observation can be explained by noting that in this low order treatment the coincidences in the case of indistinguishable input modes stem mostly from events in which two photons are present at the same input, which occurs with probability $p_0p_2+p_2p_0$. For distinguishable input modes the coincidences stem from all events that contain two photons at the input, i.e. $p_1p_1+p_0p_2+p_2p_0$. Since, according to \eqref{eq:cohfockparam}, for coherent input states, all of these probabilities scale in the same way with the mean photon number, their ratio, and thus the visibility of \eqref{eq:cohphotcoinvis}, is constant for all mean photon numbers.

\subsection{Compilation of experimental results for HOM interference at the few-photon level.}

Here we show the plots of coincidence count rates on which the few-photon values in Table~\ref{tab:visresults} of the main text are based. We restate that coincidence count rates are proportional to coincidence probabilities by a factor that is given by the average experimental repetition rate. Moreover, when calculating the HOM visibility, only the relative probabilities or count rates in a measurement are important. In the experiments we change the mutual polarization, time separation, or frequency difference of the pulses at the BS (in the main text referred to as HOM-BS) input as explained in the earlier in the Supplementary Information.\\

\underline{Deactivated memories:}
We present the data in order of the number of activated memories starting with none, i.e. pulses merely pass through attenuated to the BS. In Fig.~\ref{fig:coin-0qm}a) we show the coincidence counts as we vary the polarization difference of the pulses at the two inputs of the BS. Fitting the data to a sine function we obtain a visibility of $\mathcal{V}=50.96\pm5.56$\%. In Fig.~\ref{fig:coin-0qm}b) we display the coincidence counts as we step the temporal separation of the pulses at the two inputs of the BS. The count rates for these measurements are generally higher than all the other count rates presented. This is because this data was acquired by looking at coincidences between the transmitted part of the probe pulses in the configuration of two active quantum memories (shown in Fig.~\ref{fig:coin-2qm}b)). Hence, the balancing of the mean photon number in the transmitted pulses done less meticulously, which is the most likely reason for the observed lower visibility of $\mathcal{V}=42.43\pm2.27$\% in this case.

\begin{figure*}[t!]
	\includegraphics[width=\textwidth]{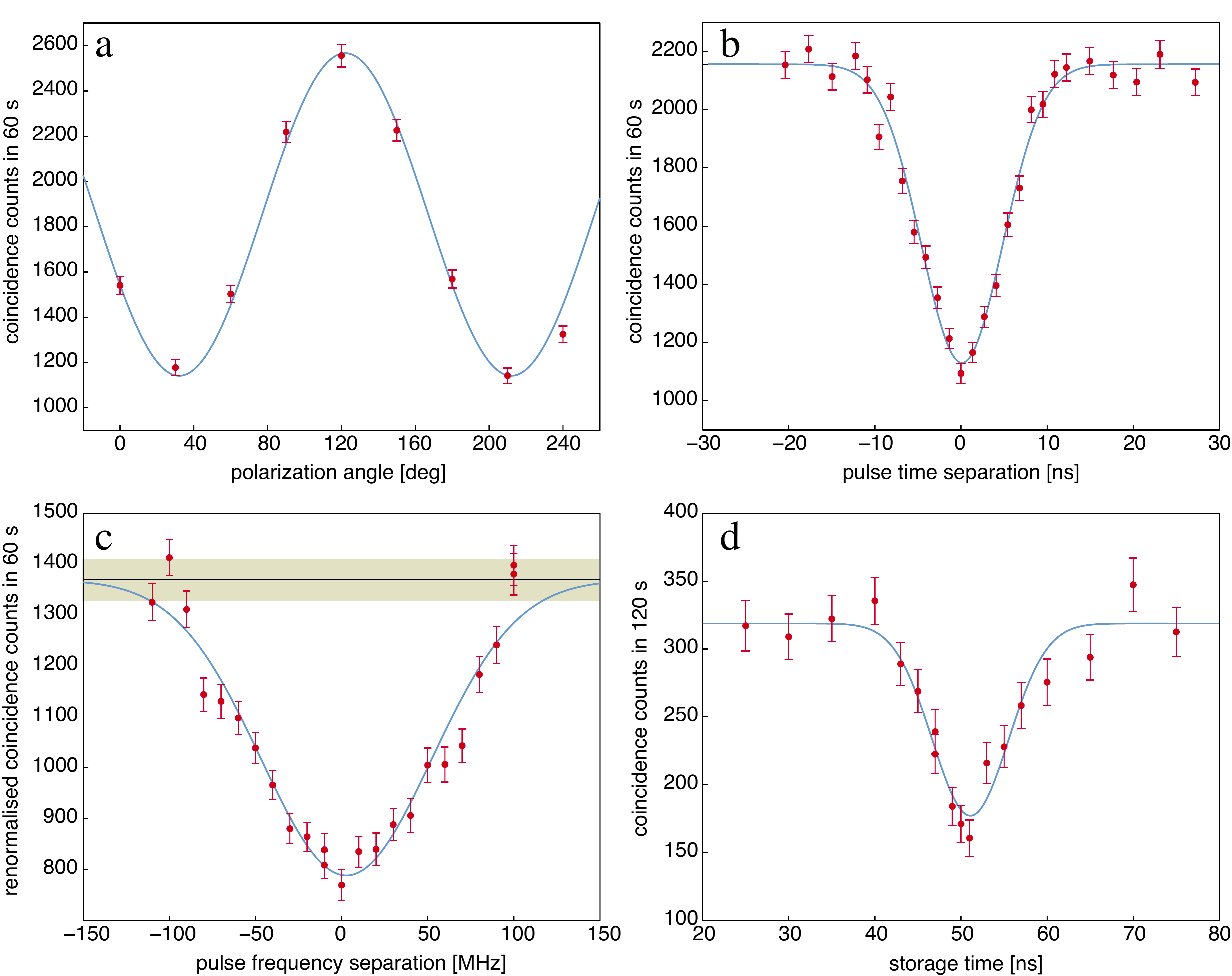}
	\caption{HOM interference manifested in coincidence counts between BS outputs with one active memory. a) Changing the polarization angle between the pulses yields a HOM visibility of $\mathcal{V}=55.51\pm4.09$\%. b) Varying the temporal overlap of the pulses produces $\mathcal{V}=47.57\pm2.96$\%. c) Altering the frequency overlap of the pulse spectra results in $\mathcal{V}=42.40\pm3.51$\%. d) Varying the storage time of the quantum memory and thus the temporal overlap of the pulses yields $\mathcal{V}=44.4\pm6.9$\%.}
	\label{fig:coin-1qm}
\end{figure*}
\begin{figure*}[t]
	\includegraphics[width=\textwidth]{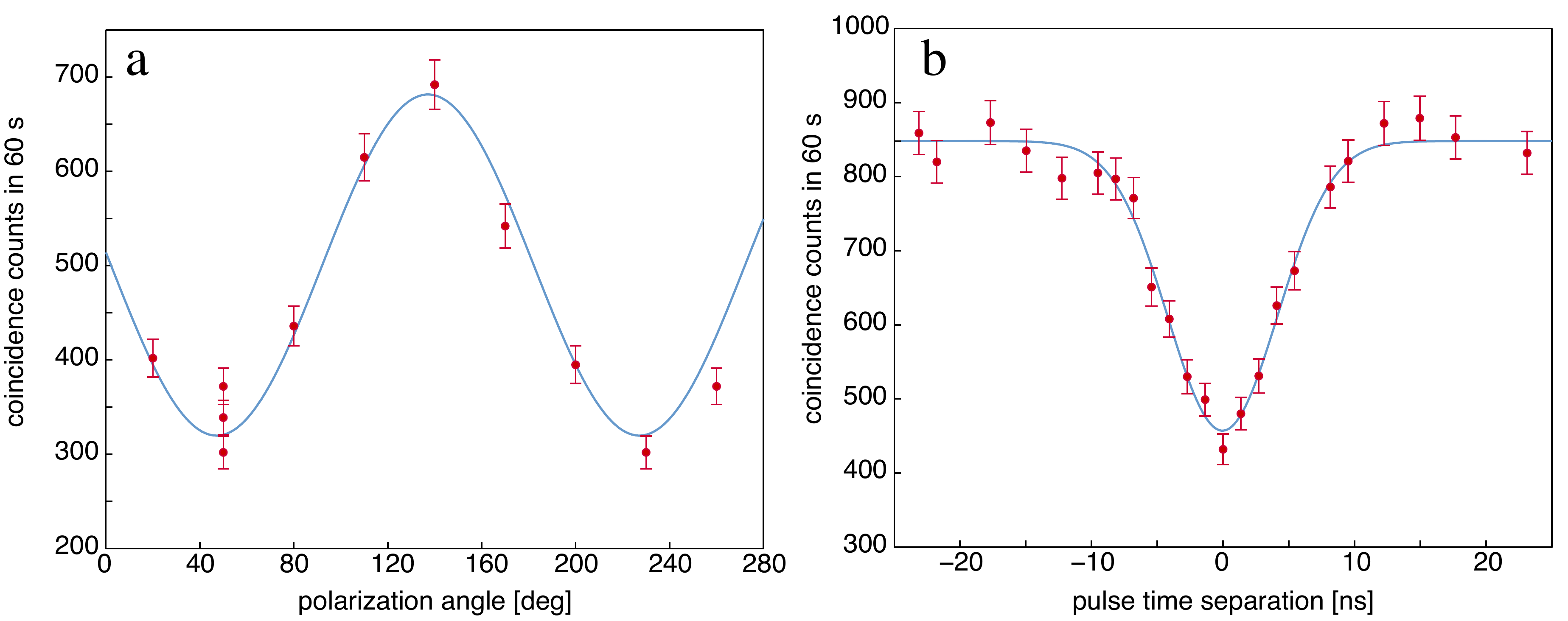}
	\caption{HOM interference manifested in coincidence counts between BS outputs with two active memories. a) Changing the polarization angle between the pulses yields a HOM visibility of $\mathcal{V}=(53.1\pm5.3)$\%. b) Varying the temporal overlap of the pulses produces $\mathcal{V}=(46.1\pm3.2)$\%.}
	\label{fig:coin-2qm}
\end{figure*}

Fig.~\ref{fig:coin-0qm}c) shows the coincidence count rates as function of the frequency difference of the two pulses at the BS inputs.
The horizontal line and surrounding shaded band shown in Fig.~\ref{fig:coin-0qm}c) -- as well as in Fig.~\ref{fig:coin-1qm}c) -- give the coincidence counts for completely distinguishable input photons as obtained by making the polarizations orthogonal. As noted earlier in the Supplementary Information, it is necessary to resort to the polarization degree of freedom in order to make the pulses completely distinguishable. The visibility from the fit is noticeably lower than that obtained when we change the other degrees of freedom. There are two main reasons for this. The first is that, in order to generate pulses with different frequencies, we drive the AOM at the limits of its bandwidth. This, in turn, necessitates setting the RF drive signal amplitude high whereby the frequency purity of the signal is contaminated by higher-order harmonics. Although it is not expected to change the maximal interference value occurring when the pulses are generated with the same modulation frequency, it will alter the shape of the interference as a function of the pulse frequency difference. Hence, the fitted Gaussian curve, assuming a Fourier limited pulse, may not correctly reproduce the actual frequency dependence of the interference. Indeed, the minimum coincidence rates consistently fall below the fitted curve. A second factor reducing the observed visibility is related to the need to adjust the AOM drive amplitude to balance the bandwidth limitation.
The limited accuracy with which we are able to estimate the appropriate RF amplitude results in significant scattering of the coincidence counts due to variations in input pulse intensities. To amend this we have found that it is necessary to normalize the points to the count rates on the individual detectors, as indicated on the y-axis of plot Fig.~\ref{fig:coin-1qm}c. Unfortunately, the manifestation of the HOM interference in the single-detector count rates -- which will be elaborated later in the Supplementary Information -- means that such a normalization procedure tends to reduce the visibility in the coincidence counts.\\

\underline{One active memory:}
Next in line are the plots for the case in which only memory $a$ is activated, while the other is left inactive. In Fig.~\ref{fig:coin-1qm} we present the coincidence count rates when changing the same degrees of freedom as in case of both memories being inactive. Additionally, in Fig.~\ref{fig:coin-1qm}d, we plot the coincidence count rates when changing the storage time in the quantum memory.\\

\underline{Two active memories:}
Lastly, we present the plots for the case in which both memories are activated. Due to limitations in our current setup it is not possible to simultaneously generate two quantum memories with different storage times, and therefore we do not acquire a storage time scan when both memories are active. Furthermore, we skip the characterization with respect to the spectral degree of freedom. The coincidence count data for the remaining two degrees of freedom are plotted in Fig.~\ref{fig:coin-2qm}, which also includes the appropriate fits.

\subsection{Manifestation of HOM interference in single detector counts.}

We also evaluate the effect of the two-photon interference on the counts registered by a single detector. This is easily done by amending the detection probability to the case of one detection event in one detector and any number of events $x$ (i.e. $x=0,1$) in the other detector. We arrive at
\begin{align}
	p_{1x}(n,m)&=1-(1-\eta_1)^n \, .
\end{align}
This expression can be inserted into \eqref{eq:nmphotbsoutdetprobcoin} to calculate $P_{1x}^{\parallel (\perp)}(n,m)$, which, through \eqref{eq:cohphotbsoutdetprobcoin}, gives us $\mathcal{P}^{\parallel (\perp)}_{1x}(\alpha,\beta)$, and from which the \emph{single-detector visibility} $\mathcal{V}_\mathrm{1x}$ is defined analogous to \eqref{eq:cohphotcoinvis}.

We can formulate a simplified expression by using the same approximations as in the case of coincidence detections:
\begin{align}
	\label{eq:cohphotbsoutdetprobsingsimp}
		\mathcal{P}^{\parallel}_{1x} & = \eta \mu + \eta\left( 2-\frac{3\eta}{4} \right)\mu^2\\ 
		\mathcal{P}^{\perp}_{1x} & = \eta \mu + \eta\left( 2-\frac{\eta}{2} \right)\mu^2 \, ,
\end{align}
from which we get the single-detector visibility
\begin{align}
	\label{eq:cohphotbsoutvissingsimp}
	\mathcal{V}_\mathrm{1x}= \frac{\eta\mu}{4+2(4-\eta)\mu} \, .
\end{align}
In the limit of low detector efficiency, $\mathcal{V}_\mathrm{1x}\approx 0$, since, in that case, the probability of detecting two photons impinging on the detector is simply twice that of detecting one. This nulls the limitation that only a single detection event can be generated per pulse. Furthermore, the single-detector visibility also goes to zero for very low mean photon numbers. In this case it is very unlikely to have two photons either at the \emph{same} or at \emph{different} input ports of the BS, hence most of the single detector counts stem from single photons from either one or the other input of the BS. It is interesting to note that if $\eta$ is known for a detector, then, from observing the single-detector visibility (see \eqref{eq:cohphotbsoutvissingsimp}), it is in principle possible to estimate the mean photon number per pulse $\mu$.
\begin{figure}[h!]
	\centering
	\includegraphics[width=\columnwidth]{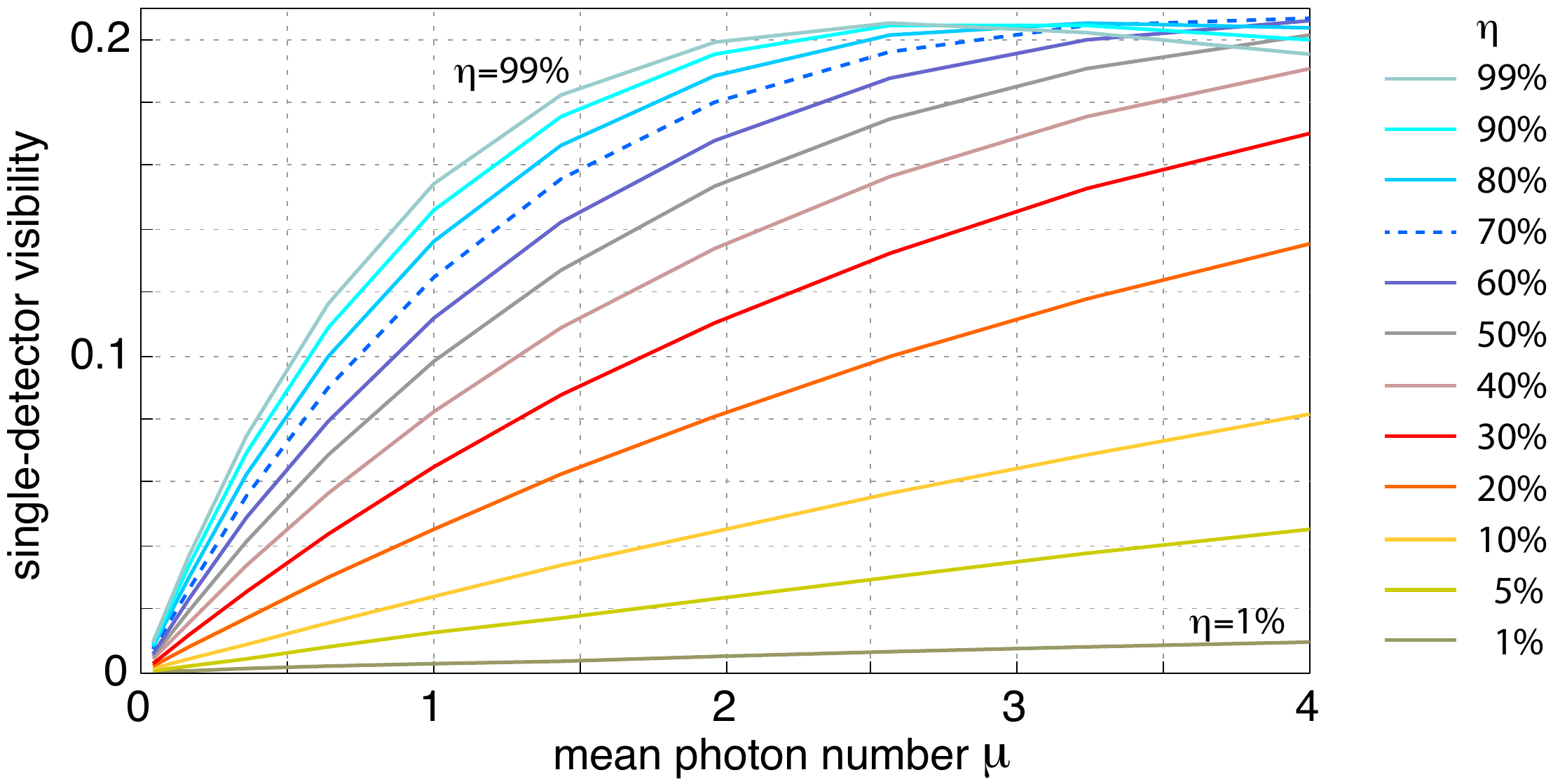}
	\caption{Plots of single-detector visibility as a function of the mean photon number for detectors with a range of single photon detection probabilities $\eta$. The $\eta=70$\% trace, highlighted with a dashed line, corresponds approximately to our detectors, which have $65\%\leq\eta\leq75\%$.}
	\label{fig:sing-theo}
\end{figure}

\begin{figure*}[t]
	\includegraphics[width=0.49\textwidth]{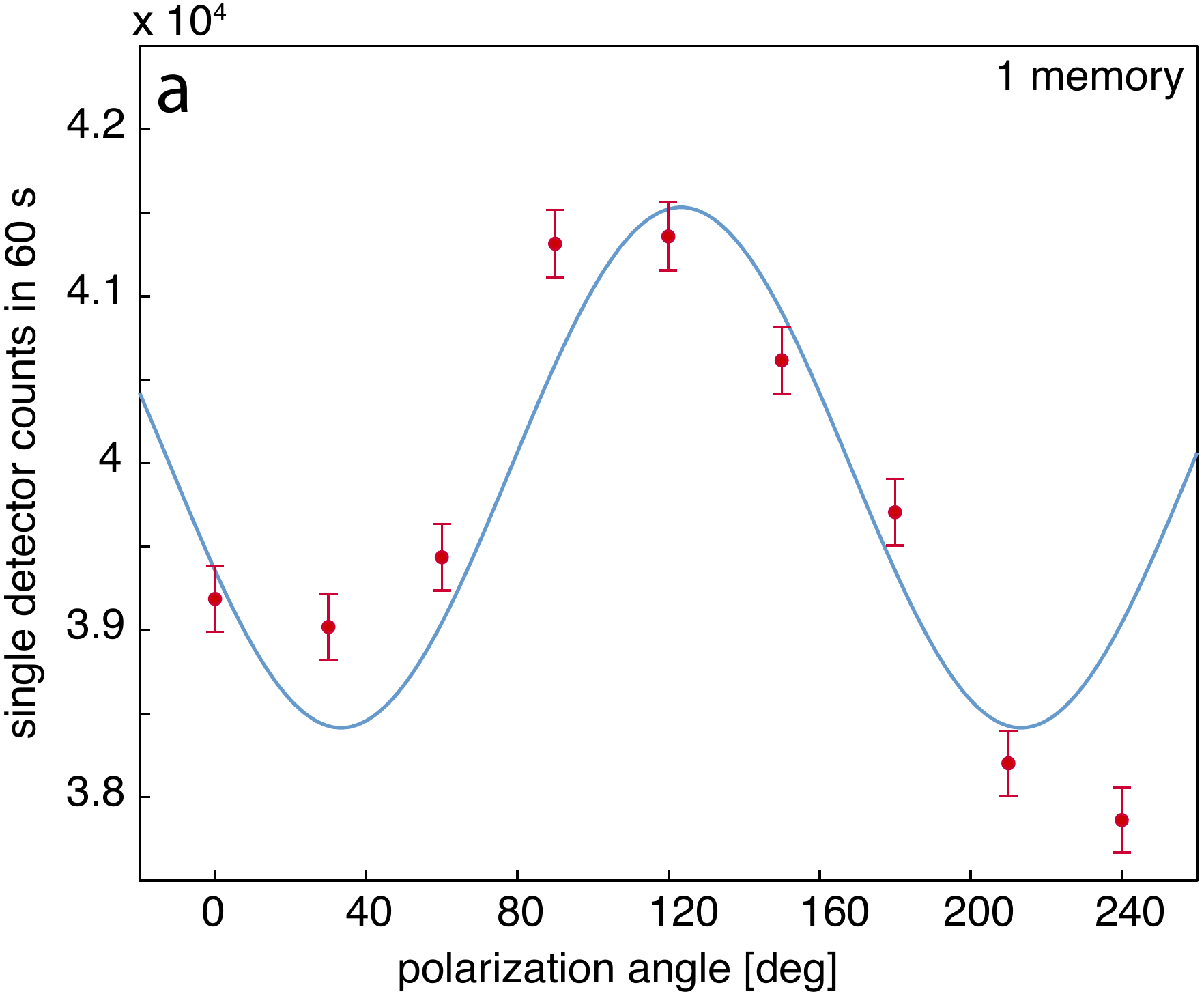}%
	\hfill%
	\includegraphics[width=0.49\textwidth]{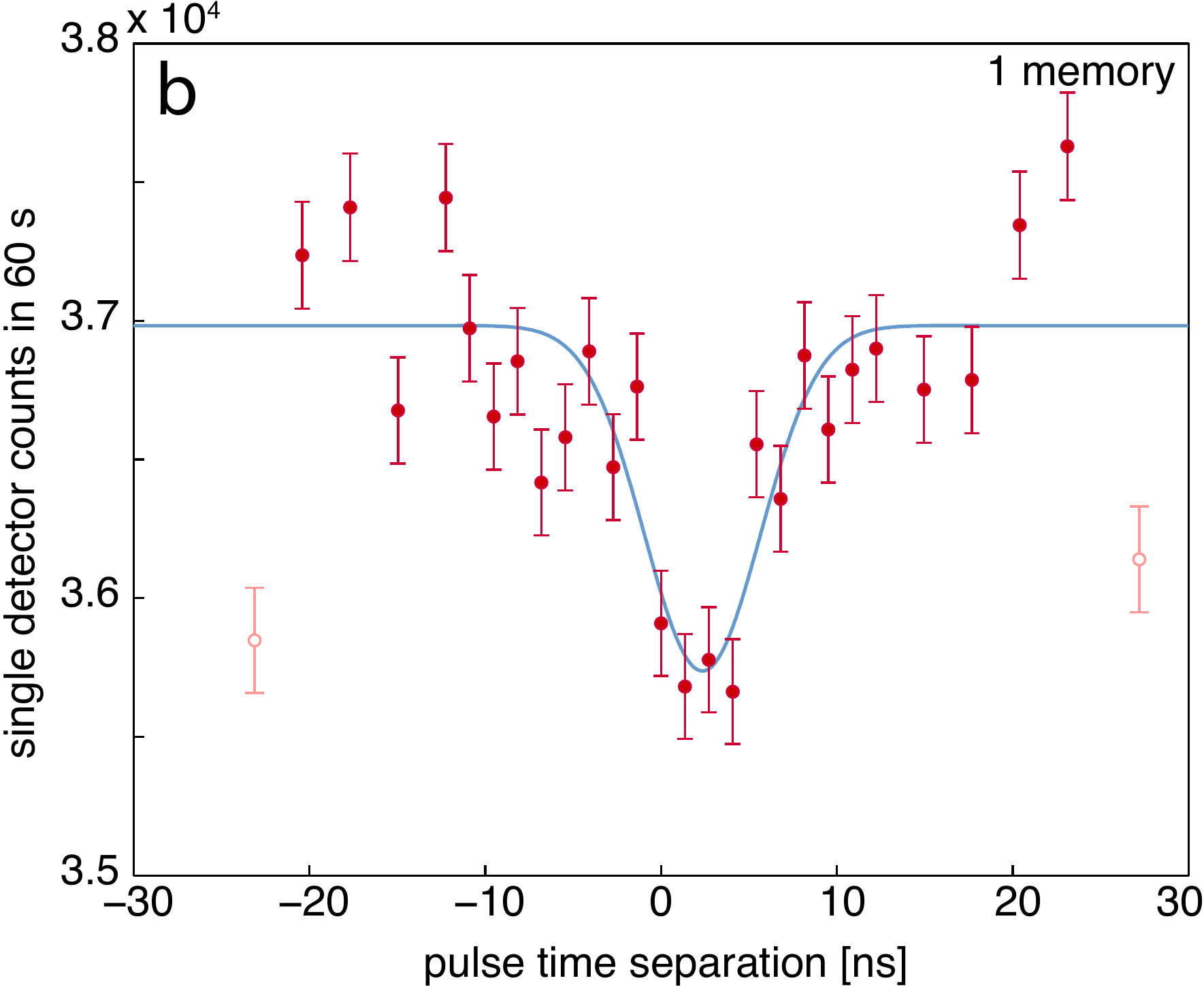}
	\caption{HOM interference manifested in single-detector counts in the case of one active quantum memory when changing a) polarization and b) time difference between pulses at BS input. For the polarization scan in a) we find $\mathcal{V}_\mathrm{1x}=(7.51\pm 3.80)$\% and for the time scan in b) we get $\mathcal{V}_\mathrm{1x}=(7.75\pm 3.25)$\%. For this measurement we only recorded the single-detector counts from Si-APD 1.}
	\label{fig:sing-1qm}
\end{figure*}
\begin{figure*}[t]
	\includegraphics[width=0.49\textwidth]{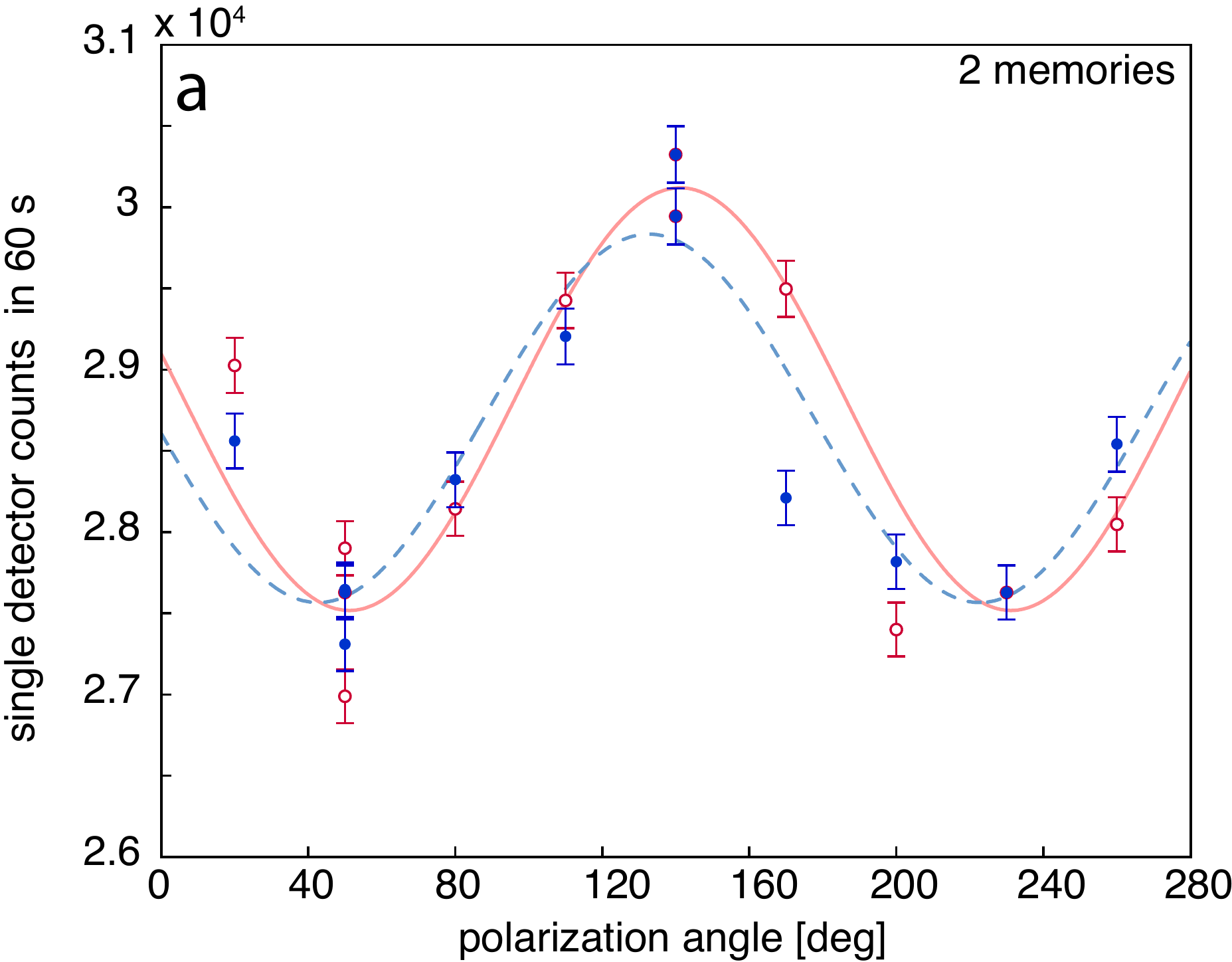}
	\hfill%
	\includegraphics[width=0.49\textwidth]{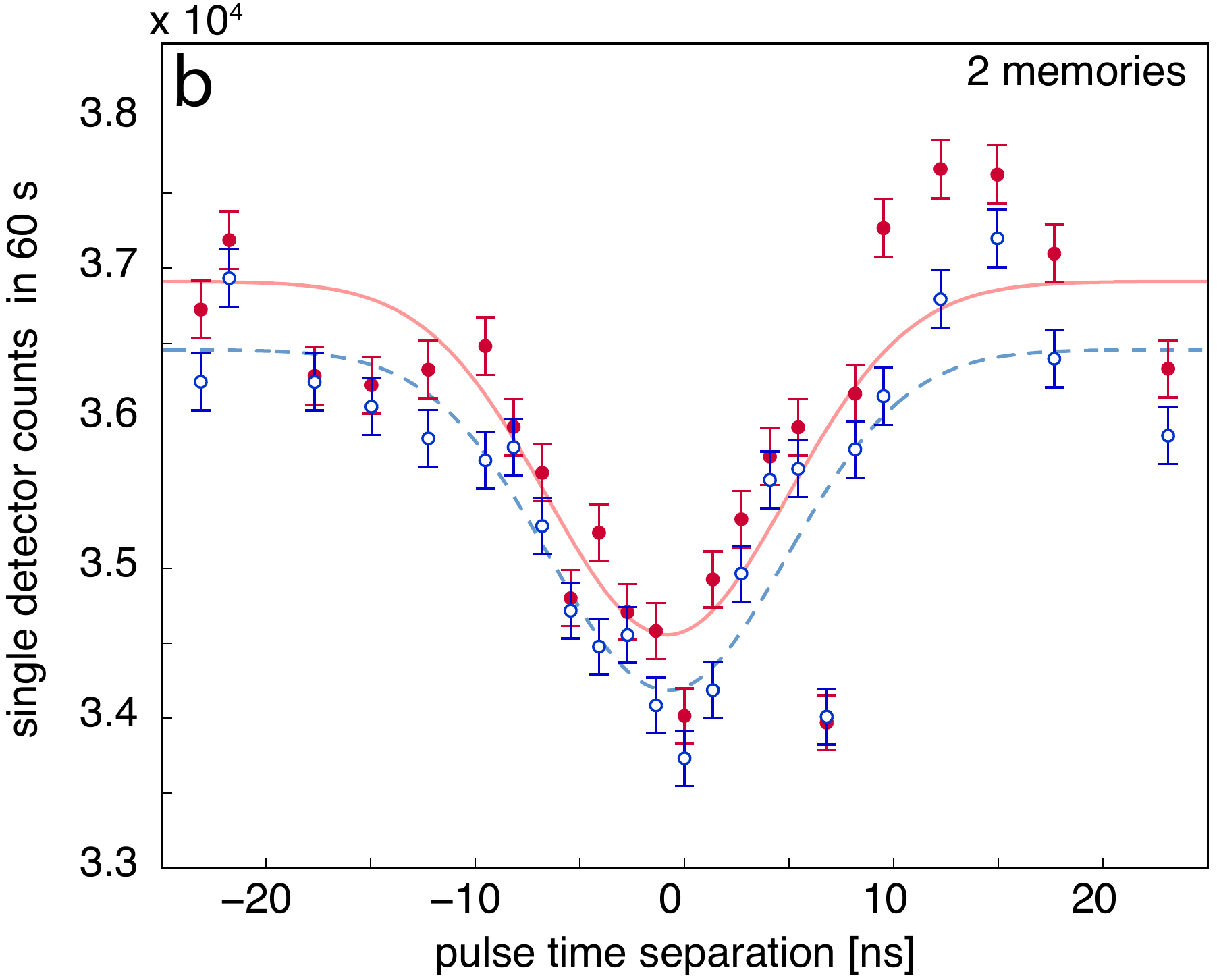}
	\caption{HOM interference manifested in single-detector counts in the case of two active quantum memories when changing a) polarization and b) time difference between pulses at BS input. For the polarization scan in a) we find $\mathcal{V}_\mathrm{1x}=(8.64\pm 2.50)$\% and $\mathcal{V}_\mathrm{1x}=(7.60\pm 2.36)$\% for Si-APD 1 and 2, respectively. For the time scan in b) we measure $\mathcal{V}_\mathrm{1x}=(6.38\pm 2.01)$\% and $\mathcal{V}_\mathrm{1x}=(6.23\pm 1.61)$\% for Si-APD 1 and 2, respectively.}
	\label{fig:sing-2qm}
\end{figure*}

Another important consequence of the manifestation of two-photon interference in the single-detector counts is that the single-detector counts cannot generally be used to normalize the coincidence counts w.r.t.~fluctuations in the input pulse intensities. Only for detectors with low detection efficiency or very low mean photon numbers, in which case $\mathcal{V}_\mathrm{1x}\approx 0$, is this normalization possible.

\subsection{Experimental results on HOM interference manifested in single-detector counts}

First, in Figure~\ref{fig:sing-1qm}, we present the single-detector counts corresponding to the coincidence counts depicted in Figure~\ref{fig:coin-1qm}a,b.
In the case where we vary the polarization and time separation we see a clear change in the single-detector counts, which, moreover, is evidently correlated with the change in coincidence counts. The count variation due to the two-photon interference is somewhat masked by the single-detector count scatter, which is due to intensity fluctuations mainly in the light going through the 10~km delay line. We fit the data in Figures~\ref{fig:sing-1qm}a and b with a sine and Gaussian function, respectively. For the former we find a mean photon number of $\mu=0.52$ while from the latter we estimate $\mu=0.54$. From the number of single-detector counts there is some evidence to conclude that the light intensity is about 15\% higher. To this should be added about 25\% uncertainty for the intensity at the BS w.r.t. the intensity at the detector due to variation in the loss in the fibre mating sleeves. Finally, the scatter of the counts makes the fits themselves rather uncertain. Nevertheless, the mere fact that the two-photon interference is manifested in the single-detector counts validates the order of magnitude of the mean photon number, as depicted in Fig.~\ref{fig:sing-theo}.

Figure~\ref{fig:sing-2qm} depicts the single-detector counts corresponding to the coincidence counts depicted in Figure~\ref{fig:coin-2qm}a,b. Again, from fitting the appropriate functions to the polarization and time data yields visibilities around 7\%, corresponding to mean photon numbers of around $\mu=0.5$.

\newpage 

\subsection{Bell-state measurement.}

In this section we derive an analytical expression for the coincidence count rates corresponding to projections onto the $|\psi^-\rangle$ Bell-state for time-bin qubits detected by the two detectors at the output of the HOM-BS. To that end, we will introduce a number of approximations as we did previously in order to calculate the HOM interference in the coincidence counts. In the limit of low mean photon numbers, two coherent states impinging onto the two inputs of a 50:50 BS can be represented in terms of Fock states as %\edit{(without normalization)}
\begin{widetext}
\begin{align}\label{eq:bsinputfockrep}
|\psi\rangle_{ab} & = \sqrt{p(1,1)}|1 1 \rangle_{a,b} + \sqrt{p(2,0)} | 2 0 \rangle_{a,b} + \sqrt{p(0,2)} | 0 2 \rangle_{a,b} \\ \nonumber 
& = \left( 
\sqrt{p(1,1)} (\hat{a}^{\dagger}\otimes \hat{b}^{\dagger})+ 
\frac{1}{\sqrt{2!}} 
\left[ \sqrt{p(2,0)} ((\hat{a}^{\dagger})^2 \otimes I)+ \sqrt{p(0,2)} (I \otimes (\hat{b}^{\dagger})^2)
\right] 
\right) |00\rangle_{a,b} \ \ , \nonumber
\end{align}
\end{widetext}
where the subscripts on the state vector refer to the order of listing the input modes, i.e. $|00\rangle_{a,b} \equiv |0\rangle_a\otimes|0\rangle_b$. The factors written as $p(n,m)$ denote the probability of having $n$ and $m$ photons in mode $a$ and $b$, and are given by $p(n,m) = |(_a\langle n|\otimes _b\langle m|)(|\alpha\rangle_a \otimes | \beta \rangle_b) |^2  = \frac{e^{-(|\alpha|^2+|\beta|^2)}}{n! m!} (|\alpha|^2)^n (|\beta|^2)^m$. Stemming from the low mean photon number assumption, we do not include terms with more than two photons. Assuming that our detectors are noiseless, terms with a total of one or no photons are also left out as they cannot generate any coincidence counts.

For a time-bin qubit, the Fock state is created in a superposition of two temporal modes, i.e., an $early$ ($e$) and a $late$ ($l$) mode, by the creation operators for the spatial input mode $x^\dagger~(x^\dagger=a^\dagger,b^\dagger)$ of the beam-splitter, as
\begin{widetext}
\begin{align}\label{eq:timebincreate}
(\hat{x}^{\dagger})^n|0\rangle_x \rightarrow \left[\cos\!\Big(\frac{\theta_x}{2}\Big)\ \hat{x}^{\dagger}_{e} \otimes I + e^{i \phi_x} \sin\!\Big(\frac{\theta_x}{2}\Big)\ I \otimes \hat{x}^{\dagger}_{l}\right]^n |00\rangle_{xe,xl} \ \ ,
\end{align}	
\end{widetext}
where $\cos\!\left(\frac{\theta_x}{2}\right)$ and $\sin\!\left(\frac{\theta_x}{2}\right)$ are the amplitudes of, and $\phi_x$ is the relative phase between, the two temporal modes composing the time-bin qubit. The subsript $xe$ refers to the early time-bin of the spatial mode $x$ and similarly for $xl$. Note, that we sometimes simplify the notation for the time-bin qubit states as $|e\rangle_x\equiv|10\rangle_{xe,xl}=(\hat{x}^{\dagger}_{e} \otimes I) |00\rangle_{xe,xl}$.
If we insert the expression in \eqref{eq:timebincreate} in place of the $\hat{a}$ and $\hat{b}$ operators in \eqref{eq:bsinputfockrep} we get the expression for the wavefunction $|\psi(\theta_a,\phi_a,\theta_b,\phi_b)\rangle_{ab}$ for time-bin qubits at the HOM-BS inputs. We split this expression into the various contributions given in \eqref{eq:bsinputfockrep}
\begin{widetext}
\begin{subequations}\label{eq:multiphotonevent}
\begin{align}
(\hat{a}^{\dagger}\otimes \hat{b}^{\dagger})|00\rangle_{ab}
&\rightarrow \frac{1}{2}\bigg[
 \left(i e^{i\phi_b} \cos\!\Big(\frac{\theta_a}{2}\Big) \sin\!\Big(\frac{\theta_b}{2}\Big) + i e^{i\phi_a} \sin\!\Big(\frac{\theta_a}{2}\Big) \cos\!\Big(\frac{\theta_b}{2}\Big) \right) \left(\hat{c}^\dagger_e\hat{c}^\dagger_l + \hat{d}^\dagger_e\hat{d}^\dagger_l\right) 
 \qquad \nonumber\\ 
&\qquad
+ \left(e^{i\phi_b} \cos\!\Big(\frac{\theta_a}{2}\Big) \sin\!\Big(\frac{\theta_b}{2}\Big) - e^{i\phi_a} \sin\!\Big(\frac{\theta_a}{2}\Big) \cos\!\Big(\frac{\theta_b}{2}\Big) \right) \left(\hat{c}^\dagger_e\hat{d}^\dagger_l - \hat{c}^\dagger_e\hat{d}^\dagger_l\right)
 \qquad \nonumber\\ 
&\qquad
+ i e^{i(\phi_a+\phi_b)}\sin\!\Big(\frac{\theta_a}{2}\Big)\sin\!\Big(\frac{\theta_b}{2}\Big)\left( (\hat{c}^\dagger_l)^2 + (\hat{d}^\dagger_l)^2 \right)
\qquad \nonumber\\ 
&\qquad
+  i \cos\!\Big(\frac{\theta_a}{2}\Big)\cos\!\Big(\frac{\theta_b}{2}\Big)\left( (\hat{c}^\dagger_e)^2 + (\hat{d}^\dagger_e)^2 \right)
\bigg] |0000\rangle_{ce,cl,de,dl} 
 \label{eq:multiphotoneventa}\qquad\\[4pt]
((\hat{a}^{\dagger})^2 \otimes I)|00\rangle_{ab} 
&\rightarrow \frac{1}{2} \bigg[ 
2e^{i\phi_a} \cos\!\Big(\frac{\theta_a}{2}\Big) \sin\!\Big(\frac{\theta_a}{2}\Big) \left(\hat{c}^\dagger_e\hat{c}^\dagger_l - \hat{d}^\dagger_e\hat{d}^\dagger_l\right)
\qquad \nonumber\\ 
&\qquad
+ i2e^{i\phi_a} \cos\!\Big(\frac{\theta_a}{2}\Big) \sin\!\Big(\frac{\theta_a}{2}\Big) \left(\hat{c}^\dagger_e\hat{d}^\dagger_l + \hat{c}^\dagger_l\hat{d}^\dagger_e\right)
\qquad \nonumber\\ 
&\qquad
+ \cos^2\!\Big(\frac{\theta_a}{2}\Big) \left((\hat{c}^\dagger_e)^2 + i2 \hat{c}^\dagger_e\hat{d}^\dagger_e
 - (\hat{d}^\dagger_e)^2 \right)
\qquad \nonumber\\ 
&\qquad
+ e^{i2\phi_a} \sin^2\!\Big(\frac{\theta_a}{2}\Big) \left((\hat{c}^\dagger_l)^2 + i2 \hat{c}^\dagger_l\hat{d}^\dagger_l
 - (\hat{d}^\dagger_l)^2 \right)
\bigg] |0000\rangle_{ce,cl,de,dl}\label{eq:multiphotoneventb}
\end{align}
\end{subequations}	
\end{widetext}
and similarly for $(I \otimes (\hat{b}^{\dagger})^2)|00\rangle_{ab}$. Again, the subscripts on the state vector refer to the order of listing the temporal and spatial modes, e.g. $ce$ labels the early bin of the spatial output mode $c$.

We will look for coincidence detection events that correspond to projections onto the Bell-state $|\psi_-\rangle_{cd} = \frac {1}{\sqrt{2}}(\hat{c}^{\dagger}_e \hat{d}^{\dagger}_l - \hat{c}^{\dagger}_l \hat{d}^{\dagger}_e) |0000\rangle_{ce,cl,de,dl}$. Such projections correspond to a detection event in the early time-bin in one detector followed by a detection event in the late time-bin in the other detector. This projection occurs with a probability $\mathcal{P}_-(\theta_a,\phi_a,\theta_b,\phi_b)=|\ _{cd}\langle\psi_- | \psi(\theta_a,\phi_a,\theta_b,\phi_b)\rangle_{cd}|^2$, which can be computed by combining \eqref{eq:multiphotonevent} with \eqref{eq:bsinputfockrep}. Assuming equal mean photon numbers at the two inputs  $|\alpha|^2=|\beta|^2\equiv\mu$ and averaging over the coherent state phases, i.e. the complex angle between $\alpha$ and $\beta$, we get the expression
\begin{widetext}
\begin{align}\label{eq:bellhomvis}
	\mathcal{P}_-(\theta_a,\phi_a,\theta_b,\phi_b)&\propto \frac{\mu^2 e^{-2 \mu}}{8} \bigg[
    4\sin^2\!\bigg(\frac{\theta_a+\theta_b}{2}\bigg) + \sin^2\!\big(\theta_a\big) + \sin^2\!\big(\theta_b\big)
	\qquad \nonumber\\ 
	&\qquad\qquad\quad
    - 2 \sin\!\big(\theta_a\big) \sin\!\big(\theta_b\big) \bigg( 1 + \cos\!\big(\phi_a-\phi_b\big) \bigg) \bigg] \ .
\end{align}	
\end{widetext}
With this we are able to calculate the probabilities of projection onto $|\psi^-\rangle$ for different combinations of qubits at the two BS inputs, i.e. for different choices of the angles $\theta_x$ and $\phi_x$. In turn, this allows us to calculate the $|\psi^-\rangle$ Bell-state measurement error rate as
\begin{align}\label{eq:errdefalt}
	e\equiv\frac{\mathcal{P}_-^{\parallel}}{\mathcal{P}_-^{\parallel}+\mathcal{P}_-^{\perp}} \, ,
\end{align}
where $\mathcal{P}_-^{\parallel}$ is the projection probability when the two input qubit states are identical, i.e. $\phi_a=\phi_b$ and $\theta_a=\theta_b$, while $\mathcal{P}_-^{\perp}$ is the projection probability for two orthogonal input qubit states. This is also defined in terms of count rates in \eqref{eq:errdef} in the main text. We will now treat a number of relevant cases.\\

%%%%%%%%%%%%%%%%%%%%%%%%%%%%%%%%%%%%%%%%%%%%%%

\underline{Expected and observed error rates when $\phi_a=\phi_b=0$.}\\ Using the simplified notation this corresponds to the case were the input qubit states are of the form $|\psi\rangle=\cos\big(\frac{\theta_x}{2}\big) |e\rangle + \sin\big(\frac{\theta_x}{2}\big) |l\rangle$.
When depicted on the Bloch sphere these qubits span the $xz$-plane. Using \eqref{eq:bellhomvis} we compute the projection probability as
\begin{widetext}
\begin{align}\label{eq:bellprojprobz}
    \mathcal{P}_-(\theta_a,0,\theta_b,0)&\propto  \frac{\mu^2 e^{-2 \mu}}{8} \bigg[
    4\sin^2\!\bigg(\frac{\theta_a+\theta_b}{2}\bigg) + \sin^2\!\big(\theta_a\big) + \sin^2\!\big(\theta_b\big)
    - 4 \sin\!\big(\theta_a\big) \sin\!\big(\theta_b\big) \bigg] \ .
\end{align}
\end{widetext}
We are interested in the probability $\mathcal{P}_-^{\parallel}$ for the case in which the input qubits are parallel ($\theta_a=\theta_b$) and $\mathcal{P}_-^{\perp}$ for the case in which the input qubit states are orthogonal ($\theta_a=\theta_b-\pi$).
Specifically, when we prepare two qubits (one at each input of the BS) in state $|e\rangle$, or two qubits in state $|l\rangle$, we expect $\mathcal{P}_-^{\parallel}=0$. The probability for observing a projection onto $|\psi\rangle$ increases as we change $\theta_a$ (or $\theta_b$), and reaches a maximum $\mathcal{P}_-^{\perp}$ if one qubit is in state $|e\rangle$ and the other one in $|l\rangle$. Hence, using the expression for the error rate above (\eqref{eq:errdefalt}), we find  $e_{e/l}^\mathrm{(att)}=0$.

We now turn to measuring the coincidence rates for all combinations of $|e\rangle$ and $|l\rangle$ input states, and thus extracting $\mathcal{P}_-^{\parallel}$ and $\mathcal{P}_-^{\perp}$, using 0.6 photons per qubit at the memory input. 
More precisely, we prepare the input qubit state $|e\rangle_{a}\otimes|e\rangle_{b}$ to measure $\mathcal{P}_-^{\parallel(1)}$ and then $|e\rangle_{a}\otimes|l\rangle_{b}$ to measure $\mathcal{P}_-^{\perp(1)}$. 
Subsequently, we prepare the input qubit state $|l\rangle_{a}\otimes|l\rangle_{b}$ to measure $\mathcal{P}_-^{\parallel(2)}$ and then $|l\rangle_{a}\otimes|e\rangle_{b}$ to measure $\mathcal{P}_-^{\perp(2)}$. These yield the average values $\mathcal{P}_-^{\parallel}=(\mathcal{P}_-^{\parallel(1)}+\mathcal{P}_-^{\parallel(2)})/2$ and $\mathcal{P}_-^{\perp}=(\mathcal{P}_-^{\perp(1)}+\mathcal{P}_-^{\perp(2)})/2$, from which we compute the experimental error rate $e_{e/l}^{(\mathrm{exp})}=0.039\pm0.037$, which is near the theoretical lowest value of $e_{e/l}^{(\mathrm{att})}=0$. \\

%%%%%%%%%%%%%%%%%%%%%%%%%%%%%%%%%%%%%%%%%%%%%%

\underline{Expected and observed error rates when $\theta_a=\theta_b=\pi/2$.} In this case the two input qubits are in equal superpositions of early and late bins, that is of the form $|\psi\rangle=\frac{1}{\sqrt{2}} \big(|e\rangle + e^{i\phi_x} |l\rangle\big)$.
On the Bloch sphere these are qubits that lie in the $xy$-plane. In this case we compute
\begin{align}\label{eq:bellprojprobx}
    \mathcal{P}_-(\pi/2,\phi_a,\pi/2,\phi_b) & \propto \frac{\mu^2 e^{-2\mu}}{4}\big(2-\cos(\phi_a-\phi_b)\big)\, ,
\end{align}
Thus the $|\psi^-\rangle$ Bell-state projection probability is smallest -- but nonzero -- when $\phi_a-\phi_b=0$, i.e. the qubit states are parallel, and largest when the phases differ by $\pi$, i.e. the qubit states are orthogonal. Inserting these values for $\mathcal{P}_-^{\parallel}$ and $\mathcal{P}_-^{\perp}$ into \eqref{eq:errdefalt} results in an expected error rate of $e_{+/-}^\mathrm{(att)}=0.25$.

Using again 0.6 photons per qubit, we measure the coincidence counts for $\phi_a-\phi_b=0$ and $\pi$ giving us $P_-^{\parallel}$ and $P_-^{\perp}$, respectively. From these we get an error rate of $e_{+/-}^\mathrm{(exp)}=0.287\pm0.020$, 
which is slightly above the theoretical bound.
This indicates that either the measurement suffers from imperfections such as detector noise or the modes at the BS are not completely indistinguishable, which in turn could be due imperfectly generated qubit states or imperfect storage of the qubit in the quantum memory. To be conservative in our assessment of our quantum memory we assume that the entire increase of the measured values of $e^{(\mathrm{exp})}$ is due to the memory fidelity being less than one.\\

%%%%%%%%%%%%%%%%%%%%%%%%%%%%%%%%%%%%%%%%%%%%%%

\underline{Bounds for attenuated laser pulses stored in quantum} \underline{and classical memories:}
We now compare the performance of our Bell-state measurement to a number of relevant bounds assuming always that any imperfections arise from the imperfect storage of the photon in the memory. 
We will derive bounds to the error rate in the case of one qubit being stored in either a classical memory (CM) or quantum memory (QM). To accommodate this scenario we assume that the memory performs the following operation $|\psi\rangle \langle \psi| \rightarrow F |\psi\rangle \langle \psi| + (1-F) |\psi^{\perp}\rangle \langle \psi^{\perp}|$, where $F$ denotes the fidelity of the stored state and $|\psi^{\perp}\rangle$ is the state orthogonal to $|\psi \rangle$. For a classical memory $F^\mathrm{CM}=0.667$~[\citenum{massar1995a}] whereas for a quantum memory $F^\mathrm{QM}=1$. 

Doing the replacement $\mathcal{P}_-^{\parallel} \rightarrow F\mathcal{P}_-^{\parallel}+(1-F)\mathcal{P}_-^{\perp}$ and likewise for $\mathcal{P}_-^{\perp}$ we can express the error rate expected after imperfect storage of one of the pulses partaking in the Bell-state measurement:
\begin{align}\label{eq:errdefaltimperf}
	e=\frac{F\mathcal{P}_-^{\parallel}+(1-F)\mathcal{P}_-^{\perp}}{\mathcal{P}_-^{\parallel}+\mathcal{P}_-^{\perp}}\, ,
\end{align}
where in this case the probabilities $\mathcal{P}_-^{\parallel}$ and $\mathcal{P}_-^{\perp}$ refer to those expected without the memory. Since the expected values for $\mathcal{P}_-^{\parallel}$ and $\mathcal{P}_-^{\perp}$ differ between the $e/l$ and $+/-$ bases we treat them separately.

Beginning with the $e/l$ basis we use \eqref{eq:errdefaltimperf} with the values from \eqref{eq:bellprojprobz} to derive a bound for the error rate of the Bell-state measurement for one of the two qubits being recalled from a quantum or a classical memory. We find that $e_{e/l}^\mathrm{(att)}=1-F$, and hence we establish the two bounds $e_{e/l}^\mathrm{(att,QM)}= 0$ and $e_{e/l}^\mathrm{(att,CM)}= 0.333$. This clearly shows that a classical memory would cause a larger error rate than the $e_{e/l}^\mathrm{(exp)}=0.039\pm0.037$ measured after storage in our memory. We can also reverse the equations and estimate our memory's fidelity based on the measured error rate. In this case, inserting $e_{e/l}^\mathrm{(exp)}$ into \eqref{eq:errdefaltimperf}, we deduce the value $F^\mathrm{exp}_{e/l}=0.961\pm0.037$.

We now turn to the $+/-$ basis. For attenuated laser pulses we insert into \eqref{eq:errdefaltimperf} the values $\mathcal{P}_-^{\parallel}=1/4$ and $\mathcal{P}_-^{\perp}=3/4$ computed from \eqref{eq:bellprojprobx}, which enables us to relate the error rate to the memory fidelity as $e_{+/-}^{}= (3-2F)/4$. Thus, one obtains the theoretical lower bound on the error rate $e_{+/-}^\mathrm{(att,QM)}=0.250$ for an ideal quantum memory $(F^\mathrm{QM}=1)$ and $e_{+/-}^\mathrm{(att,CM)}= 0.417$ with an optimal classical storage device $(F^\mathrm{CM}=2/3)$. 
We make the observation that our experimental error rate $e_{+/-}^\mathrm{(exp)}=0.287\pm0.020$ is much below the bound for a classical memory.
Based on the experimental error rate $e_{+/-}^\mathrm{(exp)}=0.287\pm0.020$ we derive an experimental value for the memory fidelity of $F^\mathrm{exp}_{+/-}=0.926\pm0.041$. The estimates of the memory fidelity $F^\mathrm{exp}_{e/l}$ and $F^\mathrm{exp}_{+/-}$ derived from our measurements in two bases are equal to within the experimental error. This together with the fact that their values are well above $0.667$ reaffirms our claim that our storage device outperforms a classical memory.

We emphasize once more that we have assumed that the reduction in error rates is due solely to the memory and thus indicates the fidelity of the memory. However, this is likely not the case as imperfections in the state preparation and detector noise also contribute to the reduction in error rate.\\

%%%%%%%%%%%%%%%%%%%%%%%%%%%%%%%%%%%%%%%%%%%%%%

\underline{Bounds for single photons stored in quantum and clas-} \underline{sical memories:} Although we do not use single photon sources for the experiments reported here, it is interesting to determine how well our results measure up to those that could have been obtained if single photon sources had been employed. In the following we will derive the error rate for the Bell-state measurement using qubits encoded into single photons. To this end we step back to \eqref{eq:bsinputfockrep}, and note that for single photon sources all probabilities are 0 except for $p(1,1)$, which describes the probability of having a single photon at each BS input. Thus, in the output state we only need to keep the terms from \eqref{eq:multiphotoneventa}, which in turn means that the Bell-state projection probability can be written as
\begin{widetext}
\begin{align}
	P_-(\theta_a,\phi_a,\theta_b,\phi_b)
	&\propto \frac{1}{4} \bigg[
    \sin^2\!\bigg(\frac{\theta_a+\theta_b}{2}\bigg) + \sin^2\!\bigg(\frac{\theta_a-\theta_b}{2}\bigg)
- \sin\!\big(\theta_a\big) \sin\!\big(\theta_b\big)\cos\!\big(\phi_a\! -\phi_b\big) \bigg] .
\end{align}
\end{widetext}
It is easily seen that for any two parallel input qubit states ($\theta_a=\theta_b$ and $\phi_a=\phi_b$) we get $P_-^\parallel=0$. Therefore, irrespective of the projection probability for orthogonal input qubit states the expected error rate is always $e^\mathrm{(sing)}=0$, where $sing$ identifies this value as belonging to the single photon case.

Gauging the effect of storing one of the single photons partaking in the Bell-state measurement in a memory is thus independent of the basis and using \eqref{eq:errdefaltimperf} we derive $e^\mathrm{(sing,QM)}=1-F^\mathrm{QM}=0$ and $e^\mathrm{(sing,CM)}=0.333$. Contrasting the error rate expected for a photon stored in a classical memory with the two values $e_{e/l}^{(\mathrm{exp})}=0.039\pm0.037$ and $e_{+/-}^\mathrm{(exp)}=0.287\pm0.020$ obtained experimentally, we recognize that both are well below $e^\mathrm{(sing,CM)}$. This means that even with a single photon source at ones disposal the error rates that we measured could not have been attained with a classical memory.\\

%%%%%%%%%%%%%%%%%%%%%%%%%%%%%%%%%%%%%%%%%%%%%%

\begin{figure}[h]
	\centering
	\includegraphics[width=\columnwidth]{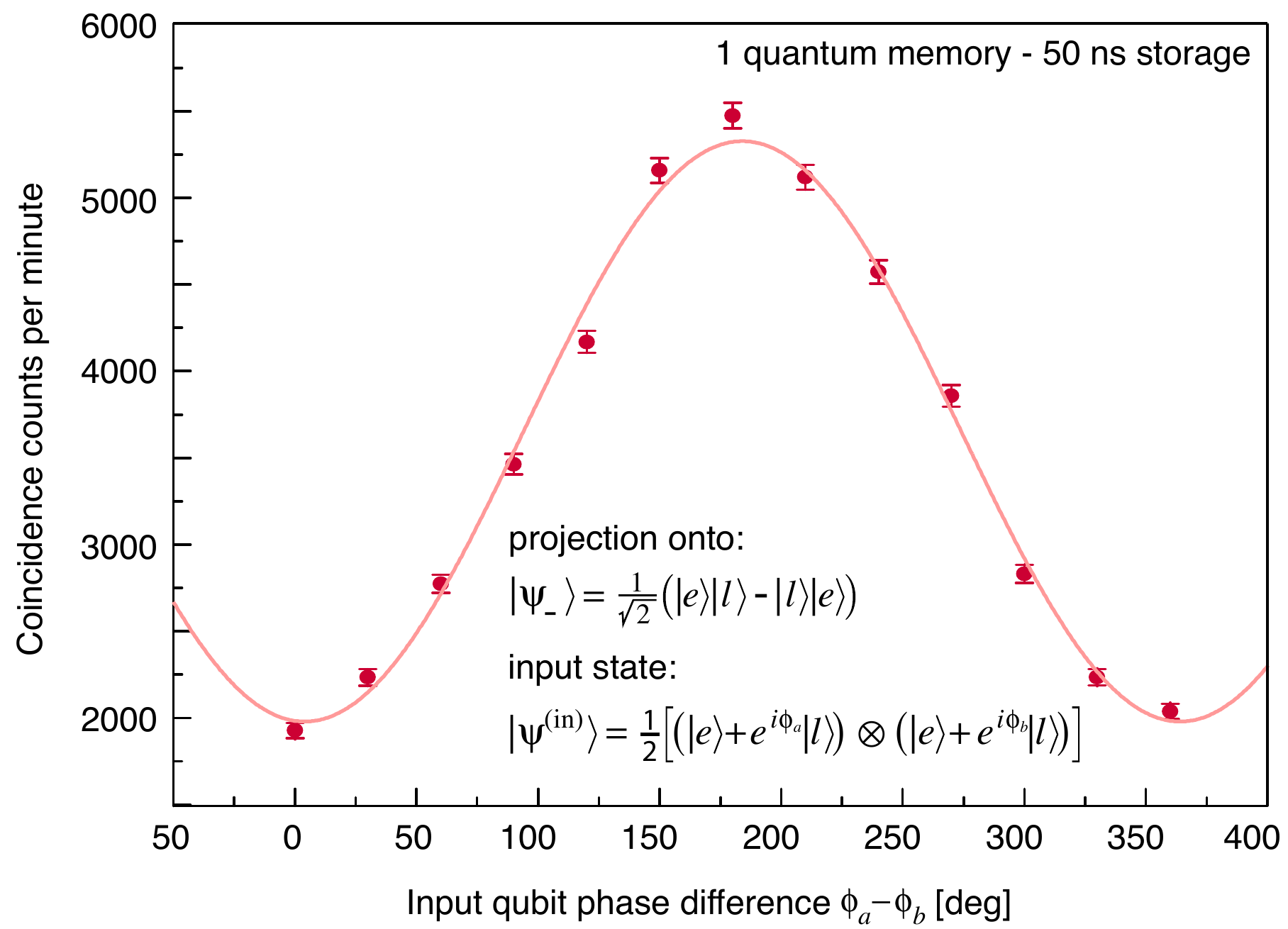}
	\caption{Rate of projection of pairs of time-bin qubits with relative phase $\phi_a-\phi_b$ onto $|\psi^-\rangle$. Each data point was acquired over 60~s}
	\label{fig:coin_1qm_tbinphase}
\end{figure}

\underline{Experiments at mean photon numbers above one.} In this final section we will explore in greater detail the HOM interference dependence on the angle $\phi_a-\phi_b$ between a set of equal superposition qubit states $|\psi\rangle_x=\frac{1}{\sqrt{2}} \big(|e\rangle + e^{i\phi_x} |l\rangle\big)$, which in line with the preceding sections belong to the $+/-$ basis. According to \eqref{eq:bellprojprobx} the coincidence count rates vary as function of $\cos\big(\phi_a-\phi_b\big)$. In Fig.~\ref{fig:coin_1qm_tbinphase} we show measured coincidence count rates as function of $\phi_a-\phi_b$ for a mean photon number per qubit before the memory of around 20.
As expected the coincidence detection probability reaches its maximum $\mathcal{P}_-^{\perp}$ when two input qubits are orthogonal ($\phi_a-\phi_b=\pi$) and when they are identical ($\phi_a-\phi_b=0$) it reaches a minimum $\mathcal{P}_-^{\parallel}$. It is natural to define a Bell-state measurement visibility as
\begin{align}\label{eq:visbsmdef}
	\mathcal{V} = \frac{\mathcal{P}_-^{\perp} - \mathcal{P}_-^{\parallel}}{\mathcal{P}_-^{\perp}}
\end{align}
analogous to \eqref{eq:HOM_Visibility} in the main text. Using values obtained from a cosine fit to the data in Fig.~\ref{fig:coin_1qm_tbinphase} yeilds $\mathcal{V}^\mathrm{exp}_\mathrm{+/-}=(62.9\pm5.2)\%$. Comparing \eqref{eq:visbsmdef} with \eqref{eq:errdefalt} it is easily seen that $\mathcal{V}$ and $e$ are related as $e_{}=(1-\mathcal{V}_\mathrm{+/-})/(2-\mathcal{V}_\mathrm{+/-})$. We can then use the expected error rates to find the corresponding Bell-state measurement visibilities. Using $e_{+/-}^{att}=0.25$ we get a theoretical value $\mathcal{V}^\mathrm{att}_\mathrm{+/-}=66.7\%$. In conclusion, our experimental Bell-state measurement visibility is only slightly below and within the experimental error actually equal to the expected value.

\subsection{References}

\hspace{1pt}\newline

\end{document}